\documentclass[12pt]{article}
\usepackage{amsmath}
\usepackage{amssymb}
\usepackage{cite}
\usepackage{float}
\usepackage{latexsym}
\usepackage{color}
\usepackage{graphicx}
\usepackage{caption}
\usepackage{hyperref}
\numberwithin{equation}{section}
\newcommand{\beq}{\begin{equation}}
\newcommand{\be}{\begin{equation}}
\newcommand{\ee}{\end{equation}}
\newcommand{\bea}{\begin{eqnarray}}
\newcommand{\eea}{\end{eqnarray}}

\newcommand{\pa}{\partial}
\newcommand{\pab}{\bar{\partial}}

\newcommand{\nn}{\nonumber}

\newcommand{\ti}[1]{\tilde{#1}}

\newcommand{\zb}{\bar{z}}
\newcommand{\tb}{\bar{t}}

\newcommand{\tp}{\tilde{\phi}}
\newcommand{\tps}{\tilde{\psi}}
\newcommand{\Mod}[1]{\ (\text{mod}\ #1)}

\begin{document}

\pagenumbering{Alph}
\begin{titlepage}
\hbox to \hsize{\hspace*{0 cm}\hbox{\tt }\hss
    \hbox{\small{\tt BRX-TH-6298}}}

\vspace{1 cm}

\centerline{\bf \Large Bosonization, cocycles, and the D1-D5 CFT }

\vspace{.6cm}

\centerline{\bf \Large on the covering surface }

\vspace{1 cm}
 \centerline{\large  Benjamin A. Burrington $^{\star}$\footnote{benjamin.a.burrington@hofstra.edu} , Amanda W. Peet$^{\dagger}$\footnote{awpeet@physics.utoronto.ca} , and Ida G. Zadeh$^{\ddagger}$\footnote{zadeh@brandeis.edu} }

\vspace{0.5cm}

\centerline{\it ${}^\star\!\!$ Department of Physics and Astronomy, Hofstra University, Hempstead, NY 11549, USA}
\centerline{\it ${}^\dagger$Department of Physics, University of Toronto, Toronto, ON M5S 1A7, Canada}
\centerline{\it ${}^\ddagger$ Martin Fisher School of Physics, Brandeis University, Waltham, MA 02454, USA}

\vspace{0.3 cm}

\begin{abstract}
We consider the D1-D5 CFT near the orbifold point, specifically the computation of correlators involving twist sector fields using covering surface techniques.  As is well known, certain twists introduce spin fields on the cover.  Here we consider the bosonization of fermions to facilitate computations involving the spin fields.  We find a set of cocycle operators that satisfy constraints coming from various $SU(2)$ symmetries, including the $SU(2)_L\times SU(2)_R$ R-symmetry.  Using these cocycles, we consider the correlator of four spin fields on the cover, and show that it is invariant under all of the $SU(2)$ symmetries of the theory.  We consider the mutual locality of operators, and compute several three-point functions.  These computations lead us to a notion of radial ordering on the cover that is inherited from the original computation before lifting. 
Further, we note that summing over orbifold images sets certain branch-cut ambiguous correlators to zero.
\end{abstract}
\end{titlepage}

\tableofcontents

\pagenumbering{arabic}
\section{Introduction}

AdS/CFT \cite{Maldacena:1997re} has provided one of the best tools for addressing questions in quantum gravity (for reviews, see \cite{Aharony:1999ti,D'Hoker:2002aw}).  In particular, the CFT description is well adapted to investigate the quantum structure of black holes, and provides some of the best evidence that their time evolution is unitary.

The D1-D5 model has proven to be particularly fertile ground for studies of black hole thermodynamics, both via direct examination of the string theory \cite{Strominger:1996sh}, and through use of AdS/CFT \cite{Maldacena:1997re}.  For this model, the near horizon of the brane configuration contains an ${\rm AdS}_3$ factor, and so the dual CFT is 2D.  The low dimension leads to simplifications and tractability in both the gravitational description \cite{Witten:1988hc,Witten:2007kt}, and in the CFT description owing to the infinite dimension of the conformal group in 2D.

An especially compelling string theoretic approach is the fuzzball programme pioneered by Mathur, reviewed in e.g. \cite{Bena:2013dka,Mathur:2011uj,Chowdhury:2010ct,Mathur:2009hf, Skenderis:2008qn,Balasubramanian:2008da,Bena:2007kg,Mathur:2005zp,Giusto:2004id,Lunin:2004uu}. A large number of microstate geometries have been generated using a variety of techniques  \cite{Bena:2015bea,Bena:2014qxa,Mathur:2012tj,Giusto:2012yz,Lunin:2012gp, Giusto:2012jx, Mathur:2011gz,Bena:2004de}. Of particular recent interest are the superstrata, which are two-parameter families of solutions rather than simple isolated solutions \cite{Bena:2015bea,Bena:2014qxa,Giusto:2012jx,Bena:2011uw}.  The orbifold CFT also appears able to capture such states \cite{Bena:2015bea}.

Work on the field theory side has been active as well \cite{Gaberdiel:2015uca,Burrington:2014yia,Carson:2014ena,Carson:2014yxa,Carson:2014xwa, Burrington:2012yq,Avery:2010vk,Avery:2010er,Avery:2010hs, Avery:2009xr, Avery:2009tu,Pakman:2009mi,Pakman:2009zz,Pakman:2009ab,Gava:2002xb,Gomis:2002qi,David:1999ec}  (for a review, see \cite{David:2002wn}).  A sizable effort has centered around deforming the theory away from the orbifold point, particularly the action on specific states \cite{Gaberdiel:2015uca,Burrington:2014yia,Carson:2014yxa,Carson:2014ena,Avery:2010vk,Avery:2010er, Avery:2010hs,Carson:2014xwa,Burrington:2012yq}. This is our main motivation for our current study as well.

The D1-D5 CFT is constructed by considering type IIB string theory compactified on $M^{4,1}\times S^1\times T^4$ (we choose $T^4$ rather than $K3$ for simplicity), with $N_1$ D1 branes wrapping the $S^1$ and $N_5$ D5 branes wrapping the $S^1\times T^4$.  The bound state of these branes, along with excitations, can be described by a field theory; considering the $T^4$ to be small, the low energy excitations will only occur along the $S^1$, and so we have a 2D CFT description.  In the gravity limit, the near horizon limit of this brane configuration is of the form $AdS_3\times S^3 \times T^4$.  In this near horizon zone, the R-symmetry is realized by the $SO(4)=SU(2)_L\times SU(2)_R$ rotations of the $S^3$, and is the $SO(4)$ that originally worked on the four Euclidean non-compact directions of $M^{4,1}$.

The moduli space of the D1-D5 CFT has been well studied \cite{Seiberg:1999xz, Giveon:1998ns, Vafa:1995bm, Larsen:1999uk, deBoer:1998ip, Dijkgraaf:1998gf,Arutyunov:1997gt, Arutyunov:1997gi, Jevicki:1998bm}.  There is believed to be a point in moduli space where the CFT becomes a free orbifold model with target space $T^4$ corresponding to a tensionless limit for the dual string theory \cite{Gaberdiel:2014cha}.  The orbifold CFT is given by $N=N_1N_5$ copies of a `base' $(c,\tilde{c})=(6,6)$ free CFT, composed of four free real scalars, four real left moving fermions, and four real right moving fermions.  The orbifold symmetry is the symmetric group $S_{N}$ which permutes copies of the CFT.

The orbifold introduces twist sectors into the theory \cite{Dixon:1986qv,Hamidi:1986vh,Dijkgraaf:1989hb}.  Twisted sector states lead to corresponding field operators that change the boundary conditions of fields near the insertion.  Correlators involving twist sector fields may be dealt with by lifting computations to a covering surface, \cite{Lunin:2000yv,Lunin:2001pw,Burrington:2012yn}, effectively `unwrapping the twist' (see also \cite{Jevicki:1998bm}).  Similar techniques are used in the condensed matter literature to compute R{\'{e}}nyi entropies, and are referred to as the `replica trick' (see \cite{Calabrese:2009qy} for a review of this technique applied to CFTs, and references therein).  R{\'{e}}nyi entropies may also be computed holographically \cite{Ryu:2006bv,Headrick:2010zt,Castro:2014tta}.

In the large $N$ limit, correlation functions that lift to a spherical cover dominate \cite{Lunin:2000yv,Lunin:2001pw}, and it is on these that we focus our attention.  Additionally, we consider the `primitive' twists corresponding to $Z_n$ elements of the full $S_N$ symmetry, use these to build full $S_{N}$ invariant objects.  Building the $S_N$ invariant objects introduces certain combinatoric factors, and these factors are exactly what lead to a suppression of factors of $1/N$ allowing one to concentrate on the case of a spherical cover \cite{Lunin:2000yv}.

The most important point for what follows is the nature of the twist fields when lifted to the covering surface.  Near the location of the twist field $z_0$, a twist field of order $n$ $\sigma_n(z_0)$ must lift to an $n$ fold copy of the plane, and so locally the map is
\be
z-z_0=a(t-t_0)^n.
\ee
We will always use $z$ to refer to the base space, and $t$ to refer to the covering space.  The map from the cover to the base is said to be ramified at the point $t_0$.  As a short hand, we will simply refer to the point as being a `ramified point' to indicate that the map has the above form near this point in the cover.  Near this point, if we consider going around the twist operator $n$ times, we take a fermion operator $\psi(z)$ to have even boundary conditions
\be
\psi\left(e^{2\pi i n}(z-z_0)\right)=\psi\left(z-z_0\right)
\ee
(we must go $n$ times around to ensure that the copy index returns to the same value).  However, this $n$-fold circling in the base space corresponds to a single circling in the cover.  Owing to the conformal weight of the fermion
\be
\psi\left(t-t_0\right)=\left(\frac{\pa z}{\pa t}\right)^{\frac{1}{2}} \psi\left((z(t)-z_0)\right)=\left(an(t-t_0)^{n-1}\right)^{\frac{1}{2}}\psi\left((z(t)-z_0)\right)
\ee
and so after the single circling in the cover,
\be
\psi\left(e^{2\pi i}(t-t_0)\right)=e^{2\pi i(n-1)/2}\left(an(t-t_0)^{n-1}\right)^{\frac{1}{2}}\psi\left(e^{2\pi i n}(z(t)-z_0)\right)=e^{2\pi i(n-1)/2}\psi\left(t-t_0\right).
\ee
Now, one can see that even though $\psi(z)$ is periodic when going around $z_0$ by circling $n$ times, owing to the conformal weight and the form of the map, $\psi(t)$ is antiperiodic when $n$ is even.  This happens for all fermions, holomorphic and antiholomorphic.  Therefore, when the twist is even, the covering surface has a spin field insertion at the ramified point \cite{Lunin:2001pw}.

Although the free orbifold theory is relatively tractable, this is not the point in the moduli space where gravitational physics may be directly addressed.  One needs to deform the theory away from the orbifold point, towards the point in moduli space with a supergravity description, to properly address burning black hole questions \cite{Avery:2010vk,Avery:2010er, Avery:2010hs,Carson:2014yxa,Carson:2014xwa, Carson:2014ena,Burrington:2014yia,Burrington:2012yq,Gaberdiel:2015uca}.  The appropriate mode for this is the deformation operator, which is in the twist-two sector of the theory \cite{David:1999ec}.  The operator product expansion (OPE) of a twist-two sector field with an odd twist will produce even twist sector fields, and the OPE of a twist-two sector field with an even twist will produce odd twist sector fields.  Hence, we expect mixing between different twist sectors.

Our main purpose for studying the spin fields closely is to be able to find just such mixing between operators, and use it to continue our earlier work to find the anomalous dimensions of candidate low-lying non-chiral fields \cite{Burrington:2012yn,Burrington:2012yq} (see also \cite{Gaberdiel:2015uca}) in conformal perturbation theory.
In our earlier work \cite{Burrington:2012yn,Burrington:2012yq}, we focused on four point correlators involving two deformation operators ${\mathcal{O}}_D$, and two copies of an operator of interest ${\mathcal{O}}_1$.  Taking the limit as one of the ${\mathcal{O}}_1$ approaches one of the ${\mathcal{O}}_D$ and inspecting the pole structure allows determination of the exchange channels, finding whether ${\mathcal{O}}_1$ mixes with other operators ${\mathcal{O}}_2$ of the same conformal dimension, and obtaining the shift to the conformal dimension \cite{Dijkgraaf:1987jt}. However, to complete this picture, the full set of ${\mathcal{O}}_2$ must be found. This is done iteratively, and for low-lying string states the set is relatively small.
%
With this information and the three-point function structure constants, the shift in conformal dimensions of the entire block can be computed.  The key point is that if ${\mathcal{O}_1}$ is of odd twist, ${\mathcal{O}_2}$ will be of even twist.  Thus, in the four point function $\langle \mathcal{O}_2 \mathcal{O}_D \mathcal{O}_D \mathcal{O}_2\rangle$ we have four even twist fields, lifting to the computation of four spin fields in the cover.  Care will be needed for such a computation, and to get the correct group theoretic structure cocycles must be used (see for example \cite{Kostelecky:1986xg} for similar considerations in the context of string theory).  It is the construction of these cocycles and verification of group theoretic constraints that occupy the remainder of this work.

The rest of the paper is organized as follows.  In section 2, we fix notation and consider group theoretic restrictions that are placed on the cocycle operators.  In section 3, we use the OPE between fermions and spin fields to determine a set of gamma matrices, and use certain freedoms in the definitions of the spin fields to fix the charge conjugation matrix to a convenient form.  We further consider the four point function between four spin fields in bosonized form, appropriately dressed with cocycles.  We determine the form of this four point function, and show that it is invariant under the symmetries of the theory, as expected.  Finally, in section 4, we consider under what circumstances the correlators are single valued.  We calculate three point correlators in the orbifold theory, and show that inheriting radial normal ordering from the base is natural in the cover, and in certain circumstances removes phase ambiguities from results.
Further, we show that summing over orbifold images sets certain ambiguous correlators to zero.
We make concluding remarks and comment on future work in section 5.
The appendices contain a number of pertinent technical details.

\section{The orbifold CFT, bosonization, and cocycles}

In this section, we fix our notation for the D1-D5 CFT near the orbifold point.  As stated above, the orbifold CFT is given by $N=N_1N_5$ copies of a (free) CFT composed of four real scalars, four real left moving fermions, and four real right moving fermions.  It is convenient to group the fermions into complex combinations.  For the left movers, we take the combinations $\psi^{\alpha \dot{A}}$ defined by (for similar notation, see \cite{Avery:2009tu})
\bea
&& \begin{pmatrix} \psi^{+{1}} \\ \psi^{-\dot{1}}\end{pmatrix}=\frac{1}{\sqrt{2}}\begin{pmatrix} \psi_1+i\psi_2 \\ \psi_3+i\psi_4 \end{pmatrix}, \\
&& \begin{pmatrix} \psi^{+{2}} \\ \psi^{-\dot{2}}\end{pmatrix}=\frac{1}{\sqrt{2}}\begin{pmatrix} \psi_3-i\psi_4 \\ -\psi_1+i\psi_2 \end{pmatrix},
\eea
and we likewise take combinations of the right movers $\ti{\psi}^{\dot{\alpha} \dot{A}}$
\bea
&& \begin{pmatrix} \ti{\psi}^{\dot{+}\dot{1}} \\ \ti{\psi}^{\dot{-}\dot{1}}\end{pmatrix}=\frac{1}{\sqrt{2}}\begin{pmatrix} \ti\psi_1+i\ti\psi_2 \\ \ti\psi_3+i\ti\psi_4 \end{pmatrix}, \\
&& \begin{pmatrix} \ti{\psi}^{\dot{+}\dot{2}} \\ \ti{\psi}^{\dot{-}\dot{2}}\end{pmatrix}=\frac{1}{\sqrt{2}}\begin{pmatrix} \ti\psi_3-i\ti\psi_4 \\ -\ti\psi_1+i\ti\psi_2 \end{pmatrix}.
\eea
The first index on the holomorphic fermion pairs, $\alpha=\{+,-\}$, is an index that measures the $SU(2)_L$ R-charge, and likewise $\dot{\alpha}=\{\dot{+},\dot{-}\}$ measures the $SU(2)_R$ R-charge for the antiholomorphic fermions.  The second index $\dot{A}=\{\dot{1},\dot{2}\}$ on the holomorphic fermion keeps track of a fake (internal) $SO(4)_I=SU(2)_1\times SU(2)_2$ charge.  We use $A$ as an index for $SU(2)_1$ and $\dot{A}$ as an index for $SU(2)_2$.  We call this a fake symmetry because it is broken by the compactification of the $T^4$ directions.  Note that the $\dot{A}$ index appears on both holomorphic and antiholomorphic fermions, i.e. they both transform under the $SU(2)_2$ part of $SO(4)_I$.

The bosons carry charges under the internal $SO_I(4)$ as well, and we write these in terms of $SU(2)_1\times SU(2)_2$ using
\be
X_{\dot{A} A}=\frac{1}{\sqrt{2}} X_\mu (\sigma^\mu)_{\dot{A} {A}}=\frac{1}{\sqrt{2}}\begin{pmatrix} X_3+iX_4 & X_1-iX_2 \\ X_1+iX_2 & -X_3+iX_4 \end{pmatrix},
\ee
where $\sigma^\mu$ are the usual Pauli matrices for $\mu=1,2,3$ and $\sigma^4=iI_{2\times2}$, and $\mu$ indices are raised and lowered with $\delta^{\mu \nu}$.

Given the above relations, there are reality conditions on the fermions and bosons
\bea
(\psi^{\alpha \dot{A}})^\dagger\equiv \psi_{\alpha \dot{A}}=-\epsilon_{\alpha \beta} \epsilon_{\dot{A}\dot{B}} \psi^{\beta \dot{B}}, \quad && \quad
(\ti{\psi}^{\dot{\alpha} \dot{A}})^\dagger\equiv \ti{\psi}_{\dot{\alpha} \dot{A}}=-\epsilon_{\dot{\alpha} \dot{\beta}} \epsilon_{\dot{A}\dot{B}} \ti{\psi}^{\dot{\beta} \dot{B}},\label{realityfermion}\\
(X_{\dot{A} {A}})^{\dagger}=\frac{1}{\sqrt{2}}\begin{pmatrix} X_3+iX_4 & X_1+iX_2 \\ X_1-iX_2 & -X_3+iX_4 \end{pmatrix} &\equiv & X^{A\dot{A}}=-\epsilon^{\dot{A}\dot{B}}\epsilon^{A B} X_{\dot{B} {B}}.\label{realityboson}
\eea
Note that the operation $\dagger$ only operates on the fields themselves, and explicit factors of $i$, not affecting `row vs. column'.  Here, and in all other places $\epsilon$ denotes the antisymmetric symbol, with $-\epsilon^{+-}=\epsilon_{+-}=1$, $-\epsilon^{\dot+\dot-}=\epsilon_{\dot+\dot-}=1$, $-\epsilon^{12}=\epsilon_{12}=1$, $-\epsilon^{\dot1\dot2}=\epsilon_{\dot1\dot2}=1$.  The OPEs for the above fields read
\bea
\psi^{\alpha \dot{A}}(z_1)\psi^{\beta \dot{B}}(z_2)\sim -\frac{1}{z_{12}}\epsilon^{\alpha \beta} \epsilon^{\dot{A}\dot{B}}+\cdots\\
\ti{\psi}^{\dot{\alpha}\dot{A}}(\zb_1)\ti{\psi}^{\dot{\beta} \dot{B}}(\zb_2)\sim -\frac{1}{\zb_{12}}\epsilon^{\dot{\alpha} \dot{\beta}} \epsilon^{\dot{A}\dot{B}}+\cdots\\
\pa X_{\dot{A}A }(z_1) \pa X_{\dot{B}B}(z_2)\sim \frac{1}{(z_{12})^2}\epsilon_{\dot{A}\dot{B}}\epsilon_{AB}+\cdots
\eea

Finally, we include a subscript $a$ on all fields that labels which of the $N=N_1N_5$ copies the field belongs to.   Thus the full set of fields is
\be
\psi^{\alpha \dot{A}}_a, \qquad \ti{\psi}^{\dot{\alpha} \dot{B}}_a, \qquad X_{a,\dot A A}.
\ee
The copy indices are permuted by members of the symmetric group $S_N$, and so the target manifold is now $(T^4)^N/S_N$.  All OPEs above simply have an $a$ and $b$ subscript appended to the two fields, and has a $\delta_{ab}$ appended to the right hand side.

The orbifold introduces new twisted sectors into the theory.  As mentioned in the introduction, we will always deal with twist operators by lifting the problem to a covering surface \cite{Lunin:2000yv,Lunin:2001pw}, and focus on the case where the covering surface is a sphere.  Of particular interest for us will be the case of even twists, where in the cover an insertion of a spin field must be made.

One way of dealing with spin fields is to write them explicitly in terms of bosonized fermions, see for example \cite{Kostelecky:1986xg}.  One copy of the CFT is all that will appear in computations on the covering surface, and so we focus on this case.

\subsection{Bosonization and fermions}

In this section, we deal with the bosonization of the fermions.  Henceforth, we will assume that we are working on the covering space, and so we use $t$ coordinates to denote the location of operators.

Following the conventions of \cite{Lunin:2001pw}, we will use the indices $5,6$ to label the two holomorphic bosons whose exponentials are the holomorphic fermions, and we will use the indices $\ti{5}, \ti{6}$ to refer to their antiholomorphic counterparts.  Cocycles will be written as dressing operators in the following way:
\be
e^{i\pi c_{\vec{k}}}e^{i \vec{k}\cdot\vec{\phi}},
\ee
where $\vec{\phi}$ refers to a vector of scalar fields that are the bosonized fermions from both the left and right moving sectors, i.e. $\vec{k}\cdot \vec{\phi}= k_{5}\phi^5+k_{6}\phi^6+k_{\ti 5}\ti{\phi}^{\ti 5}+k_{\ti 6}\ti{\phi}^{\ti 6}$.  The cocycle operator is taken to be linear in the momenta $\vec{k}$ as well as being linear in the momentum operators $\vec{\alpha}_0$, and so we write
\be
c_{\vec{k}}= \vec{k}\cdot\left(M\vec{\alpha}_0\right)=k^i M_{ij}\alpha_0^j.
\ee
Here, $M$ is an arbitrary matrix of constants that we will put restrictions on shortly.  We use the vector notation to refer to both left and right moving momenta together.  If we need to refer to only the holomorphic part, we will use $\vec{k}_L$ and $\vec{k}_R$ to refer to the left and right moving momenta individually ($\alpha$s will in addition have tildes to denote right-movers).  It is the commutation relations between $\vec{\alpha}_0$ and the fields $\vec{\phi}$,
\be
[\phi^{i},\alpha^j_{0}]=i\delta^{i j},
\ee
that lead to phases giving proper commutation relations amongst the fermions.  Finally, the cocycle operators introduce multiple modes in multiple exponentials at the same point to construct an operator.  Hence, taking the conjugate is now performed by
\be
\left(e^{i\pi c_{\vec{k}}} e^{i \vec{k}\cdot \vec{\phi}}\right)^\dagger=e^{-i \vec{k}\cdot \vec{\phi}}e^{-i\pi c_{\vec{k}}}=e^{[-i \vec{k}\cdot \vec{\phi}, -i\pi c_{\vec{k}}]}e^{-i\pi c_{\vec{k}}}e^{-i \vec{k}\cdot \vec{\phi}}=e^{-i\pi k^i M_{ij} k^j}e^{-i\pi c_{\vec{k}}}e^{-i \vec{k}\cdot \vec{\phi}}. \label{conjrule}
\ee

For the system at hand, we have $4$ holomorphic fermions, with a reality constraint.  We take
\bea
\psi^{+ \dot{1}}=e^{-i\pi M_{6i}\alpha_0^i}e^{-i \phi^6},\label{ferp1}\\
\psi^{+ \dot{2}}=e^{i\pi M_{5i}\alpha_0^i}e^{i \phi^5}.\label{ferp2}
\eea
The reality condition \ref{realityfermion} along with \ref{conjrule} gives
\bea
\psi^{-\dot{2}}=-e^{-i\pi M_{66}}e^{i\pi M_{6i}\alpha_0^i}e^{i\phi^6},\label{ferm2}\\
\psi^{-\dot{1}}=e^{-i\pi M_{55}}e^{-i\pi M_{5i}\alpha_0^i}e^{-i\phi^5}.\label{ferm1}
\eea
Similarly for the antiholomorphic fermions we find
\bea
&&\tps^{\dot{+}\dot{1}}=e^{-i\pi M_{\ti{6} i}\alpha^i_0}e^{-i\tp^{\ti{6}}},\\
&&\tps^{\dot{+}\dot{2}}=e^{i\pi M_{\ti{5} i}\alpha_0^i}e^{i\tp^{\ti 5}},\\
&&\tps^{\dot{-} \dot{1}}=e^{-i\pi M_{\ti{5}\ti{5}}}e^{-i\pi M_{\ti{5}i}\alpha_0^i}e^{-i \tp^{\ti 5}},\\
&&\tps^{\dot{-}\dot{2}}=-e^{-i\pi M_{\ti{6} \ti{6}}}e^{i\pi M_{\ti{6}i}\alpha_0^i}e^{i\tp^{\ti{6}}}.
\eea

Next, we need to guarantee that all of the fermions anticommute.  This is guaranteed for fermions constructed from the same bosonized operator because the OPE introduces factors of $t_{12}$ which are antisymmetric under interchange of $1\leftrightarrow 2$.  This is just the usual way that a single complex fermion $\psi$ anticommutes with itself and with its conjugate $\psi^\dagger$.  However, we must use the cocycles to guarantee that the fermions constructed using different bosonic fields also anticommute.

To check this, we compare the products of $\psi^{+\dot{2}}(t_1)\psi^{-\dot{2}}(t_2)$ and $\psi^{-\dot{2}}(t_2)\psi^{+\dot{2}}(t_1)$ to make sure these are antisymmetric.  Using (\ref{ferp2}) and (\ref{ferm2}) we find for the first ordering
\bea
\psi^{+\dot{2}}(t_1)\psi^{-\dot{2}}(t_2)&=&e^{i\pi M_{5i}\alpha_0^i}e^{i\phi^{5}(t_1)}\left(-e^{-i\pi M_{66}}e^{i\pi M_{6 i}\alpha_0^i} e^{i\phi^6(t_2)}\right) \\
&=&-e^{-i\pi M_{66}}e^{i\pi \left(M_{5i}+M_{6i}\right)\alpha_0^i}e^{[i\phi^5,i\pi M_{6i}\alpha_0^i]} e^{i\phi^5(t_1)} e^{i\phi^6(t_2)}\nn\\
&=&-e^{-i\pi M_{66}}e^{i\pi \left(M_{5i}+M_{6i}\right)\alpha_0^i}e^{-i\pi M_{6 5}} e^{i\phi^5(t_1)} e^{i\phi^6(t_2)}.\nn
\eea
In the last line above, it does not matter in which order we write the last two exponentials of fields: they do not share an OPE and so they commute.  We compare this to the second ordering
\bea
\psi^{-\dot{2}}(t_2)\psi^{+\dot{2}}(t_1)&=&-e^{-i\pi M_{66}}e^{i\pi M_{6 i}\alpha_0^i} e^{i\phi^6(t_2)}e^{i\pi M_{5i}\alpha_0^i}e^{i\phi^{5}(t_1)} \\
&=&-e^{-i\pi M_{66}}e^{i\pi \left(M_{5i}+M_{6i}\right)\alpha_0^i}e^{-i\pi M_{5 6}} e^{i\phi^5(t_2)} e^{i\phi^6(t_1)}.\nn
\eea
These two expressions must be negatives of each other.  Taking the ratio, and requiring this to be $-1$ we find
\be
e^{i\pi(M_{56}-M_{65})}=-1
\ee
Thus, we conclude that
\be
M_{56}-M_{65}\equiv 1 \Mod{2} \label{firstmod}
\ee
The usual choices of $\pm 1$ could work in the above, but the congruency is all that is necessary.  This immediately suggests introducing the antisymmetric part of the matrix $M$, and we define
\be
A_{ij}=M_{ij}-M_{ji}
\ee
where $a,b$ are indices that run over the bosonized fermion indices $5,6,\ti{5},\ti{6}$.
Equation (\ref{firstmod}) quickly generalizes for the other fermion combinations as well.  So, we find
\bea
A_{ij}\equiv 1 \Mod{2}. \label{amod2const}
\eea

\subsection{Symmetry generators}

Next, we consider various $SU(2)$ symmetry currents in the theory.  We have the R-symmetry for the holomorphic sector ($SU(2)_L$)
\bea
&& J^+=\psi^{+\dot{1}}\psi^{+\dot{2},}\\
&& J^-=-\psi^{-\dot{1}}\psi^{-\dot{2}},\\
&& J^3=-\frac{1}{2}\left(\psi^{+\dot{1}}  \psi^{-\dot{2}}-\psi^{+\dot{2}}\psi^{-\dot{1}}\right),
\eea
and the R-symmetry for the antiholomorphic sector ($SU(2)_R$)
\bea
&&
\ti{J}^{\dot{+}}=\tps^{\dot{+}\dot{1}}\tps^{\dot{+}\dot{2}},\\
&& \ti{J}^{\dot{-}}=-\tps^{\dot{-}\dot{1}}\tps^{\dot{-}\dot{2}},\\
&& \ti{J}^{\dot{3}}=-\frac{1}{2}\left(\tps^{\dot{+}\dot{1}} \tps^{\dot{-}\dot{2}}-\tps^{\dot{+}\dot{2}}\tps^{\dot{-}\dot{1}}\right).
\eea
In addition to these, we have the $SU(2)_2$ from the $SO(4)=SU(2)_1 \times SU(2)_2$ symmetry which is generated on the holomorphic side by
\bea
&& {\mathcal J}^{\uparrow}=\psi^{+\dot{1}}\psi^{-\dot{1}},\\
&& {\mathcal J}^{\downarrow}=-\psi^{+\dot{2}}\psi^{-\dot{2}},\\
&& {\mathcal J}^0=-\frac{1}{2}\left(\psi^{+\dot{1}}\psi^{-\dot{2}}-\psi^{-\dot{1}}\psi^{+\dot{2}}\right)= -\frac{1}{2}\left(\psi^{+\dot{1}}\psi^{-\dot{2}}+\psi^{+\dot{2}}\psi^{-\dot{1}}\right),
\eea
and on the antiholomorphic side
\bea
&& \ti{{\mathcal J}}^{\uparrow}=\tps^{\dot{+}\dot{1}}\tps^{\dot{-}\dot{1}},\\
&& \ti{{\mathcal J}}^{\downarrow}=-\tps^{\dot{+}\dot{2}}\tps^{\dot{-}\dot{2}},\\
&& \ti{{\mathcal J}}^0=-\frac{1}{2}\left(\tps^{\dot{+}\dot{1}}\tps^{\dot{-}\dot{2}}- \tps^{\dot{-}\dot{1}}\tps^{\dot{+}\dot{2}}\right)= -\frac{1}{2}\left(\tps^{\dot{+}\dot{1}}\psi^{-\dot{2}}+\tps^{\dot{+}\dot{2}}\psi^{-\dot{1}}\right).\label{Jcal0}
\eea

Of course the $SU(2)_2$ is really generated by the sum of the generators, because this is a symmetry that affects both the holomorphic and antiholomorphic fields:
\be
{\mathbb J}^i= {\mathcal J}^i + \ti{\mathcal J}^i.
\ee
This is still incomplete because the bosons also transform under this symmetry; however, we will not have use of these here.  For the time being, we will treat the two halves of this $SU(2)$ operating on the fermions as being independent.

Next, we need to express these symmetry generators in bosonic form.  We will go through only one of these exercises: the others follow from similar calculations.  We first write
\be
J^+=\psi^{+\dot{1}}\psi^{+\dot{2}}=e^{-i\pi M_{6i}\alpha_0^i}e^{-i\phi^6}e^{i\pi M_{5i}\alpha_0^i}e^{i\phi^5},
\ee
where we have have used equations (\ref{ferp1}) and (\ref{ferp2}). In the above, we do not need to worry about the normal ordering of the fields: they share no OPE, and so are automatically normal ordered.  Moving the cocycles to the left, we find
\be
J^+=e^{i\pi M_{56}}e^{i\pi\left(M_{5i}-M_{6i}\right)\alpha_0^i}e^{i(\phi^5-\phi^6)}. \label{bosj+}
\ee
Similarly, we compute
\bea
J^-
&=& e^{i\pi\left(-M_{55}-M_{66}+M_{65}\right)} e^{i\pi\left(-M_{5i}+M_{6i}\right)\alpha_0^i}e^{i\left(-\phi^5+\phi^6\right)},\label{bosj-}
\eea
where we have used equations (\ref{ferm2}) and (\ref{ferm1}). Next, to compute $J^3$, we need to keep track of the normal ordering.  In this case, we simply need to subtract the divergent simple pole of the form $1/(t_1-t_2)$. We then find
\be
J^3
=\frac{i}{2}(\pa\phi^5(t_2)-\pa\phi^6(t_2)), \label{bosj3}
\ee
and so no additional phases are introduced for the operator $J^3$.  A similar calculation holds for the $\ti{J}^i, {\mathcal J}^i$ and $\ti{\mathcal J}^i$ operators.  The bosonized form of these generators are listed in appendix \ref{app_Js}.

\subsection{Spin fields, OPEs, and more constraints}\label{spinfields}

Recall from the introduction that even twists introduce spin fields on the cover.  Here, we will identify the spin fields corresponding to Ramond ground states, and label them according to their $SU(2)$ symmetry transformations.  We start with the family that has $S^{+\dot{+}}$ as the top component.  To start, we define
\be
S^{+\dot{+}}=e^{i\theta_{pp}}e^{\frac{i\pi}{2}\left(M_{5i}-M_{6i}+M_{\ti{5}i}-M_{\ti{6i}}\right)\alpha_0^i}
e^{\frac{i}{2}(\phi^5-\phi^6+\tp^{\ti{5}}-\tp^{\ti{6}})} \label{s++ini}
\ee
where $\theta_{pp}$ is an arbitrary phase that we will take advantage of later.  To this spin field, we apply $SU(2)$ lowering operators to find the other members of this multiplet.  We start with
\bea
S^{-\dot{+}}\equiv[J_0^-,S^{+\dot{+}}]=Res_{t\rightarrow 0} J^{-}(t)S^{+\dot{+}}(0,0).
\eea
We will do this first computation in full detail, and simply state the results for other computations. We have, using (\ref{s++ini}) and (\ref{bosj-}),
\bea\label{Smp_detail}
S^{-\dot{+}}&=&Res_{t\rightarrow 0} e^{i\pi(-M_{55}-M_{66}+M_{65})}e^{i\pi(-M_{5i}+M_{6i})\alpha_0^i}e^{i(-\phi^5(t)+\phi^6(t))} \label{s-+ini}\\
&&\qquad \qquad \times e^{i\theta_{pp}}e^{\frac{i\pi}{2}\left(M_{5i}-M_{6i}+M_{\ti{5}i}-M_{\ti{6i}}\right)\alpha_0^i}e^{\frac{i}{2}(\phi^5(0)-\phi^6(0)+\tp^{\ti{5}}(0)-\tp^{\ti{6}}(0))}\nn\\
&=&e^{i\theta_{pp}}e^{\frac{i\pi}{2}\Big(
\begin{smallmatrix}
-M_{55}& +M_{65}& +M_{\ti{5}5}& - M_{\ti{6}5} \\
-M_{56} & -M_{66} & -M_{\ti{5}6} & +M_{\ti{6}6}\end{smallmatrix}\Big)} e^{\frac{i\pi}{2}\left(-M_{5i}+M_{6i}+M_{\ti{5}i}-M_{\ti{6}i}\right)\alpha_0^i}
e^{\frac{i}{2}\left(-\phi^5+\phi^6+\tp^{\ti{5}}-\tp^{\ti{6}}\right)},\nn
\eea
where we have opted to write the overall phase in two lines: this is \emph{not} meant as a matrix, and is rather just a sum of terms leading to an overall phase (but written in the above way to organize the terms). Similarly, we compute $S^{+\dot{-}}\equiv[\ti{J}_0^{\dot{-}},S^{+\dot{+}}]=Res_{\tb\rightarrow 0} \ti{J}^{\dot{-}}(\tb)S^{+\dot{+}}(0,0)$ to find
\be
S^{+\dot{-}}=
    e^{i\theta_{pp}}e^{\frac{i\pi}{2}\Big(
    \begin{smallmatrix}
    M_{5\ti{5}}& -M_{6\ti{5}}& -M_{\ti{5}\ti{5}}& + M_{\ti{6}\ti{5}} \\
    -M_{5\ti{6}} & +M_{6\ti{6}} & -M_{\ti{5}\ti{6}} & -M_{\ti{6}\ti{6}} \end{smallmatrix}\Big)} e^{\frac{i\pi}{2}\left(M_{5i}-M_{6i}-M_{\ti{5}i}+M_{\ti{6}i}\right)\alpha_0^i}
e^{\frac{i}{2}\left(\phi^5-\phi^6-\tp^{\ti{5}}+\tp^{\ti{6}}\right)}. \label{s+-ini}
\ee

Next, we have two ways to compute $S^{-\dot{-}}$.  We may either compute $[\ti{J}^{\dot{-}},S^{-\dot{+}}]$ or we may compute $[J^{-},S^{+\dot{-}}]$.  Doing both of these calculations we find
\bea
&& \kern-4em [\ti{J}^{\dot{-}},S^{-\dot{+}}]=e^{i\theta_{pp}}e^{\frac{i\pi}{2}\left(\begin{smallmatrix}
    -M_{55} & + M_{65} & + M_{\ti{5} 5} & - M_{\ti{6} 5} \\
    -M_{56} & - M_{66} & - M_{\ti{5} 6} & + M_{\ti{6} 6} \\
    -M_{5\ti{5}} & + M_{6\ti{5}} & - M_{\ti{5} \ti{5}} & + M_{\ti{6} \ti{5}} \\
    +M_{5\ti{6}} & - M_{6\ti{6}} & - M_{\ti{5} \ti{6}} & - M_{\ti{6} \ti{6}} \\
    \end{smallmatrix}
    \right)}e^{\frac{i\pi}{2}(-M_{5i}+M_{6i}-M_{\ti{5}i}+M_{\ti{6}i})\alpha_0^i} e^{\frac{i}{2}(-\phi^5+\phi^6-\tp^{\ti{5}}+\tp^{\ti{6}})}, \label{s--ini}\\
&& \kern-4em  [{J}^{{-}},S^{+\dot{-}}]=e^{i\theta_{pp}}e^{\frac{i\pi}{2}\left(\begin{smallmatrix}
    -M_{55} & + M_{65} & - M_{\ti{5} 5} & + M_{\ti{6} 5} \\
    -M_{56} & - M_{66} & + M_{\ti{5} 6} & - M_{\ti{6} 6} \\
    +M_{5\ti{5}} & - M_{6\ti{5}} & - M_{\ti{5} \ti{5}} & + M_{\ti{6} \ti{5}} \\
    -M_{5\ti{6}} & + M_{6\ti{6}} & - M_{\ti{5} \ti{6}} & - M_{\ti{6} \ti{6}} \\
    \end{smallmatrix}
    \right)}e^{\frac{i\pi}{2}(-M_{5i}+M_{6i}-M_{\ti{5}i}+M_{\ti{6}i})\alpha_0^i} e^{\frac{i}{2}(-\phi^5+\phi^6-\tp^{\ti{5}}+\tp^{\ti{6}})},
\eea
where again, we write the overall phase in four lines to help organize terms, and is not meant as a matrix.  The two dressing phases differ by a phase
\be
e^{i\pi(M_{\ti{5}5}-M_{5\ti{5}})}e^{i\pi(M_{5\ti{6}}-M_{\ti{6}5})} e^{i\pi(M_{6\ti{5}}-M_{\ti{5}6})}e^{i\pi(M_{\ti{6}6}-M_{6\ti{6}})},
\ee
which is identically +1, given the earlier constraints (\ref{amod2const}) on the antisymmetric part of $M$.  In fact, one can see this in a simple way.  The two orders of applying $J^-$ and $\ti{J}^{\dot{-}}$ are related by moving the currents past each other.  This results in moving a two-fermion object past another two-fermion object.  This should result in four minus ones coming up in the computation.  These are exactly the four terms above.  This is not surprising: the algebra of moving $J$ and $\ti{J}$ past each other only depends on getting the bosonized fermions to anticommute, and so the above offers only a check of earlier computations.

With this family of spin fields, i.e. $S^{+\dot{+}}$, $S^{+\dot{-}}$, $S^{-\dot{+}}$, and $S^{-\dot{-}}$, we may come up with our next non trivial condition on the matrix $M$.  This comes about because we expect that
\be
S^{\alpha \dot{\alpha}}(t_1,\tb_1)S^{\beta \dot{\beta}}(t_2,\tb_2)\sim\frac{K \epsilon^{\alpha \beta}\epsilon^{\dot{\alpha}\dot{\beta}}}{|t_1-t_2|}+\cdots, \label{OPEexpected}
\ee
where $K$ is a complex normalization constant.  Given this expectation, the OPE of $S^{+ \dot{+}}(t_1,\tb_1)$ with $S^{- \dot{-}}(t_2,\tb_2)$ should have the same phase as $S^{- \dot{-}}(t_1,\tb_1)S^{+ \dot{+}}(t_2,\tb_2)$.  This is guaranteed because
\be
e^{i\pi c_{\vec{k}}}e^{i\vec{k}\cdot\vec{\phi}}(t_1,\tb_1)e^{i\pi c_{-\vec{k}}}e^{-i\vec{k}\cdot\vec{\phi}}(t_2,\tb_2)=e^{[i\vec{k}\cdot\vec{\phi},i\pi c_{-\vec{k}}]}e^{i\vec{k}\cdot\vec{\phi}}(t_1,\tb_1) e^{-i\vec{k}\cdot\vec{\phi}}(t_2,\tb_2),
\ee
and
\be
e^{i\pi c_{-\vec{k}}}e^{-i\vec{k}\cdot\vec{\phi}}(t_1,\tb_1)e^{i\pi c_{\vec{k}}}e^{i\vec{k}\cdot\vec{\phi}}(t_2,\tb_2)=e^{[-i\vec{k}\cdot\vec{\phi},i\pi c_{\vec{k}}]}e^{-i\vec{k}\cdot\vec{\phi}}(t_1,\tb_1) e^{i\vec{k}\cdot\vec{\phi}}(t_2,\tb_2).
\ee
The leading order singularity on the right hand side is given\footnote{We emphasize here that the $\vec{k}\cdot\vec{k}=k^{5}k^{5}+k^{6}k^{6}+k^{\ti 5}k^{\ti 5}+k^{\ti 6}k^{\ti 6}$, i.e. it is a sum over both left and right moving momenta separately.} by $1/|t_1-t_2|^{\vec{k}^2}$  coming from contracting the exponentials.  Therefore, the phases only differ by the commutator terms.  However, these commutator terms are identical
\be
e^{[-i\vec{k}\cdot\vec{\phi},i\pi c_{\vec{k}}]}=e^{[i\vec{k}\cdot\vec{\phi},i\pi c_{-\vec{k}}]}
\ee
because $c_{-\vec{k}}=-c_{\vec{k}}$ (the cocycles are linear in the momentum).  Thus, the leading order term in the OPE is symmetric.  This is, in fact, the main problem when considering only the holomorphic side by itself: the OPE was guaranteed to be symmetric, and could not accommodate the antisymmetry of a single $\epsilon^{\alpha \beta}$ on the right hand side.  Further, the right hand side would be ambiguous as to the phase because of the branch cut in a term $1/(t_1-t_2)^{\frac12}$.  In our OPE above, the branch cuts from $1/(t_1-t_2)^{\frac12}$ and $1/(\tb_1-\tb_2)^{\frac12}$ cancel, yielding unambiguous results.  This is the main reason that we treat \emph{both} the holomorphic and antiholomorphic parts of the spin fields simultaneously.

However, this tells us that the leading order term in the OPE of $S^{+ \dot{-}}(t_1,\tb_1)S^{- \dot{+}}(t_2,\tb_2)$ must be the negative of the leading order term of the $S^{+ \dot{+}}(t_1,\tb_1)S^{- \dot{-}}(t_2,\tb_2)$ term.  This single sign will guarantee the above $ \epsilon^{\alpha \beta}\epsilon^{\dot{\alpha}\dot{\beta}}$ structure appearing on the right hand side of (\ref{OPEexpected}).  We compute only these two OPEs, and the others are guaranteed by the properties of the general OPE.

So, we need to compute the OPE $S^{+\dot{+}}(t_1,\tb_1)S^{-\dot{-}}(t_2,\tb_2)$ and $S^{+\dot{-}}(t_1,\tb_1)S^{-\dot{+}}(t_2,\tb_2)$ and compare the phases of the leading order singularity.  We start with the $S^{+\dot{+}}S^{-\dot{-}}$ OPE, and using (\ref{s++ini}) and (\ref{s--ini}) we find
\be
S^{+\dot{+}}(t_1,\tb_1)S^{-\dot{-}}(t_2,\tb_2)=
e^{2i\theta_{pp}}e^{\frac{i\pi}{4}\left(\begin{smallmatrix}
-M_{55} & + M_{65} & + 3M_{\ti{5} 5} & - 3M_{\ti{6} 5} \\
-3M_{56} & - M_{66} & - 3M_{\ti{5} 6} & + 3M_{\ti{6} 6} \\
-M_{5\ti{5}} & + M_{6\ti{5}} & - M_{\ti{5} \ti{5}} & + M_{\ti{6} \ti{5}} \\
+M_{5\ti{6}} & - M_{6\ti{6}} & - 3M_{\ti{5} \ti{6}} & - M_{\ti{6} \ti{6}} \\
\end{smallmatrix}
\right)} \times\left( \frac{1}{|t_1-t_2|}+...\right).
\ee
Similarly, we use (\ref{s+-ini}) and (\ref{s-+ini}), to compute
\be
S^{+\dot{-}}(t_1,\tb_1)S^{-\dot{+}}(t_2,\tb_2)=
e^{2i\theta_{pp}}e^{\frac{i\pi}{4}\left(\begin{smallmatrix}
-M_{55} & + M_{65} & + M_{\ti{5} 5} & - M_{\ti{6} 5} \\
-3M_{56} & - M_{66} & - M_{\ti{5} 6} & + M_{\ti{6} 6} \\
+M_{5\ti{5}} & - M_{6\ti{5}} & - M_{\ti{5} \ti{5}} & + M_{\ti{6} \ti{5}} \\
-M_{5\ti{6}} & + M_{6\ti{6}} & - 3M_{\ti{5} \ti{6}} & - M_{\ti{6} \ti{6}} \\
\end{smallmatrix}
\right)} \times\left( \frac{1}{|t_1-t_2|}+...\right).
\ee
The difference in the phases must be $-1$.  This gives the requirement
\be
e^{\frac{i\pi}{4}\left(\begin{smallmatrix}
2(M_{\ti{5}5} -M_{5\ti{5}})& +2(M_{5\ti{6}}-M_{\ti{6}5})\\ +2(M_{6\ti{5}}-M_{\ti{5}6}) &+ 2(M_{\ti{6}6}-M_{6\ti{6}}) \end{smallmatrix}\right)}=-1,
\ee
which yields the condition
\bea
&& -A_{5\ti{5}}+ A_{5\ti{6}}+ A_{{6\ti{5}}}- A_{6\ti{6}}\equiv 2 \Mod{4}, \label{sectppcond} \label{ppsector}
\eea
where again $A_{ij}=M_{ij}-M_{ji}$.  Note that both holomorphic and antiholomorphic fields need to be considered simultaneously to arrive at this; each $A$ has one index from the holomorphic fields, and one from the antiholomorphic fields.  This is clearly consistent with the earlier constraints.  For example, one solution to the above is $A_{5 \ti{5}}=-1,A_{5\ti{6}}=1, A_{6\ti{5}}=1, A_{6 \ti{6}}=1$.

The bosonized form of the other three families of spin fields are listed in appendix \ref{app_spinfields}. We can then look at similar calculations as those leading to (\ref{sectppcond}) in the other sectors.  These give three more constraints
\bea
A_{5\ti{5}} +A_{5\ti{6}}-A_{6\ti{5}}-A_{6\ti{6}}\equiv 2\; \Mod{4} \label{p1sector},\\
A_{5\ti{5}} -A_{5\ti{6}}+A_{6\ti{5}}-A_{6\ti{6}}\equiv 2\; \Mod{4} \label{1psector},\\
-A_{5\ti{5}} -A_{5\ti{6}}-A_{6\ti{5}}-A_{6\ti{6}}\equiv 2\; \Mod{4} \label{11sector}.
\eea%

These are not independent constraints given the earlier constraints $A_{ij}\equiv 1 \Mod{2}$ (\ref{amod2const}).  This relation gives that $A_{ij}\equiv \pm 1 \Mod{4}$.  These combine into the relation $A_{ij}\pm 2\equiv -A_{ij} \Mod{4}$.  Thus, the three equations (\ref{p1sector})-(\ref{11sector}) are obtained by adding 4 to both sides of equation (\ref{sectppcond}), and splitting this 4 into a 2 added to one of the $A_{ij}$ and a second 2 added to another $A_{ij}$, flipping two signs on the left hand side of (\ref{ppsector}), arriving at (\ref{p1sector})-(\ref{11sector}).

There are many solutions to the above constraints.  We have already identified the solution
\bea
(A_{5\ti{5}},A_{5\ti{6}},A_{6\ti{5}},A_{6\ti{6}})=(-1,1,1,1).
\eea
We can, of course, generate more.  For example, we can add 4 to any of the above $A_{ij}$ entries and still match the constraints. Similarly for the already mentioned shifts of the form $(A_{5\ti{5}},A_{5\ti{6}},A_{6\ti{5}},A_{6\ti{6}})\rightarrow (A_{5\ti{5}}\pm 2,A_{5\ti{6}}\pm2,A_{6\ti{5}},A_{6\ti{6}})$.  This adding or subtracting $2$ does not affect the $\!\Mod{2}$ congruences.  In the $\!\Mod{4}$ congruences, the two $A$ components that we have shifted either have the same sign, or different sign in any given constraint.  If the resulting 2s in the constraints cancel, then the constraint is left unaltered.  If the 2s add up, the result is a $\pm 4$, which does not affect a congruency mod 4.

Finally, constraining the $SS$ OPE to have the correct symmetry yields a charge conjugation matrix $C$ (moderately generalized) so that we may write
\be
\left(S^{W\ti{X}}\right)^\dagger\equiv S_{W\ti{X}}=C_{W\ti{X} Y\ti{Z} }S^{Y\ti{Z}},
\ee
where the double index notation is understood to be the natural matrix product for the outer product of the left and right $SO(4)$ spin representations (these $SO(4)$ groups are constructed from $SU(2)_2\times SU(2)_L$ and $SU(2)_2\times SU(2)_R$ respectively, and simply rotate the four real fermions on the left or the right together).  We will discuss this natural matrix multiplication in more detail in the next section.

\section{Symmetry, OPEs, and correlators on the cover}
\subsection{Fermion/spin-field OPEs and gamma matrices}

We expect the leading order OPE between a fermion and spin field to furnish a set of gamma matrices.  Let us begin by considering the fermion $\psi^{-\dot{1}}(t)$.  This field acts as a lowering operator for the left moving R-symmetry, and a raising operator for the left moving part of the $SU(2)_2$, and so to give a nonzero answer, it must either work on $S^{+ \ti{X}}$ or $S^{\dot{2} \ti{X}}$.  Further, because the index $X$ is one of $4$ right moving labels $\dot{+}, \dot{-}, \dot{1}, \dot{2}$, we have 8 calculations to perform for this fermion.  We perform 2 explicitly here, and simply write down the results for the other computations.

So, starting with (\ref{ferm1}) and (\ref{s++ini}), we find
\bea
&&\psi^{-\dot{1}}(t) S^{+\dot{+}}(0,0)=\\
&&\qquad e^{-i\pi M_{55}}e^{-i\pi M_{5i}\alpha_0^{i}} e^{-i\phi^5}(t)\,e^{i\theta_{pp}}e^{\frac{i\pi}{2}(M_5i-M_6i+M_{\ti{5} i}-M_{\ti{6}i})\alpha_0^i} e^{\frac{i}{2}(\phi^5-\phi^6+\tp^{\ti{5}}-\tp^{\ti{6}})}(0,0).\nn
\eea
We follow our usual prescription of moving cocycles to the left.  This will bring the two operators containing $\phi$ fields next to each other, which we then expand using the general rule for OPEs of exponentials.  We find
\bea
{\psi^{-\dot{1}}(t) S^{+\dot{+}}(0,0)}&=&e^{i\theta_{pp}}e^{-i\pi M_{55}}e^{-i\pi M_{5i}\alpha_0^{i}} e^{[-i\phi^5,\frac{i\pi}{2}(M_{5i}-M_{6i}+M_{\ti{5} i}-M_{\ti{6}i})\alpha_0^i]}\\
&& \times  e^{\frac{i\pi}{2}(M_{5i}-M_{6i}+M_{\ti{5} i}-M_{\ti{6}i})\alpha_0^i} e^{-i\phi^5}(t) e^{\frac{i}{2}(\phi^5-\phi^6+\tp^{\ti{5}}-\tp^{\ti{6}})}(0,0)\nn\\
&=&e^{i\theta_{pp}} e^{-i\pi M_{55}}e^{\frac{i\pi}{2}(M_{55}-M_{65}+M_{\ti{5} 5}-M_{\ti{6}5})} \nn \\
&&\times e^{\frac{i\pi}{2}(-M_{5i}-M_{6i}+M_{\ti{5} i}-M_{\ti{6}i})\alpha_0^i} e^{\frac{i}{2}(-\phi^5-\phi^6+\tp^{\ti{5}}-\tp^{\ti{6}})}(0,0)\frac{1}{t^{\frac{1}{2}}}\left(1+{\mathcal O}(t)\right)\nn.
\eea
We see that the remaining operator at $(0,0)$ is a spin field, and is in fact $S^{\dot{1}\dot{+}}$, up to the factor of $e^{i\theta_{1p}}$; see appendix \ref{app_spinfields}.  We then find
\be
\psi^{-\dot{1}}(t) S^{+\dot{+}}(0,0)=e^{i(\theta_{pp}-\theta_{1p})}e^{\frac{i\pi}{2}(-M_{55}-M_{65}+M_{\ti{5}5} -M_{\ti{6} 5})} S^{\dot{1}\dot{+}}(0,0)\frac{1}{t^{\frac12}}\left(1+{\mathcal O}(t)\right).\label{firstgammacalc}
\ee
Thus, we have identified one of the non-zero components of the gamma matrix associated with the zero mode of $\psi^{-\dot{1}}$.  Application of $\psi^{-\dot{1}}$ to $S^{+\dot-}$ follows from the right moving $SU(2)_R$ symmetry and yields the same phase factor as in (\ref{firstgammacalc}), as can be checked explicitly.


In total, we have eight fermions (four right moving, and four left moving), with eight nonzero phases associated with each one.  This leads to 64 phases to calculate, which can be grouped into gamma matrices to efficiently express the fermion-spin field OPES as
\bea
\psi^{\alpha \dot{A}}(t)S^{W\ti{X}}(0,0)=(\Gamma^{\alpha \dot{A}})^{W\ti{X}}\,_{Y\ti{Z}} S^{Y\ti{Z}}(0,0)\frac{1}{t^{\frac{1}{2}}}(1+{\mathcal{O}}(t)), \\
\tilde{\psi}^{\dot{\alpha} \dot{A}}(\tb)S^{W\ti{X}}(0,0)= (\ti{\Gamma}^{\dot{\alpha}\dot{A}})^{W\ti{X}}\,_{Y\ti{Z}} S^{Y\ti{Z}}(0,0)\frac{1}{\tb^{\frac{1}{2}}}(1+{\mathcal{O}}(\tb)),
\eea
where capital letters near the end of the alphabet run over $+,-,\dot{1},\dot{2}$, and capital letters near the end of the alphabet with a tilde run over $\dot{+},\dot{-},\dot{1},\dot{2}$.  This gives a set of $16\times 16$ gamma matrices.  Matrix multiplication is defined in terms of summation over pairs of indices, hence
\be
(AB)^{U\ti{V}}\,_{Y\ti{Z}}=A^{U\ti{V}}\,_{W\ti{X}}B^{W\ti{X}}\,_{Y\ti{Z}}.
\ee
We collect the nonzero parts of the gamma matrices in appendix \ref{gammaappendix}.

Recall that $\psi^{-\dot{1}}$ is conjugate to $\psi^{+\dot{2}}$.  Therefore, we can form hermitian gamma matrices from
\be\label{gamma34}
\Gamma^{3}= \left(\Gamma^{-\dot{1}}+\Gamma^{+\dot{2}}\right),\qquad\Gamma^{4}= \frac{1}{i}\left(\Gamma^{-\dot{1}}-\Gamma^{+\dot{2}}\right).
\ee
However, $\psi^{+\dot{1}}$ and $\psi^{-\dot{2}}$ are anti-conjugate, owing to the epsilon structure in the reality conditions (\ref{realityfermion}).  Therefore, the hermitian combinations in this case are
\be\label{gamma12}
\Gamma^{1}=\left(\Gamma^{+\dot{1}}-\Gamma^{-\dot{2}}\right),\qquad\Gamma^{2}=\frac{1}{i}\left(\Gamma^{+\dot{1}}+\Gamma^{-\dot{2}}\right).
\ee
Similarly for the antiholomorphic side, we define
\bea\label{gammat1234}
&& \ti{\Gamma}^{\ti{1}}= \left(\ti{\Gamma}^{\dot{+}\dot{1}}-\ti{\Gamma}^{\dot{-}\dot{2}}\right),\qquad
\ti{\Gamma}^{\ti{2}}=\frac{1}{i}\left(\ti{\Gamma}^{\dot{+}\dot{1}} +\ti{\Gamma}^{\dot{-}\dot{2}}\right),\\
&& \ti{\Gamma}^{\ti{3}}= \left(\ti{\Gamma}^{\dot{-}\dot{1}}+\ti{\Gamma}^{\dot{+}\dot{2}}\right),\qquad
\ti{\Gamma}^{\ti{4}}= \frac{1}{i}\left(\ti{\Gamma}^{\dot{-}\dot{1}}-\ti{\Gamma}^{\dot{+}\dot{2}}\right).\nn
\eea
These gamma matrices satisfy
\be
\left\{\Gamma^\mu,\Gamma^\nu \right\}=2\delta^{\mu\nu}, \quad  \left\{\ti{\Gamma}^{\ti{\mu}},\ti{\Gamma}^{\ti{\nu}} \right\}=2\delta^{\ti{\mu}\ti{\nu}}, \quad \left\{\Gamma^\mu,\ti{\Gamma}^{\ti{\nu}}\right\}=0. \label{cliffalg}
\ee

We define the left moving chirality matrix
\be
\Gamma^5 \equiv \Gamma^1\Gamma^2\Gamma^3\Gamma^4 =\begin{pmatrix} 1_{8\times8} & 0 \\ 0 & -1_{8\times8}\end{pmatrix}=\gamma^5\otimes 1
\ee
and the right moving chirality matrix
\bea
\ti{\Gamma}^{\ti{5}}&=& \ti{\Gamma}^{\ti{1}}\ti{\Gamma}^{\ti{2}}\ti{\Gamma}^{\ti{3}}\ti{\Gamma}^{\ti{4}}  \\
&=&
\begin{pmatrix}
1_{2\times 2} & 0 & 0 & 0 & 0 & 0 & 0 & 0 \\
0 & -1_{2\times 2} & 0 & 0 & 0 & 0 & 0 & 0  \\
0 & 0 & 1_{2\times 2} & 0 & 0 & 0 & 0 & 0 \\
0 & 0 & 0 & -1_{2\times 2} & 0 & 0 & 0 & 0 \\
0 & 0 & 0 & 0 & 1_{2\times 2} & 0 & 0 & 0 \\
0 & 0 & 0 & 0 & 0 & -1_{2\times 2} & 0 & 0  \\
0 & 0 & 0 & 0 & 0 & 0 & 1_{2\times 2} & 0  \\
0 & 0 & 0 & 0 & 0 & 0 & 0 & -1_{2\times 2} \\
\end{pmatrix}=1\otimes\gamma^5\nn
\eea
with
\be
\gamma^5=\begin{pmatrix} 1_{2\times 2} & 0 \\ 0 & -1_{2\times 2} \end{pmatrix},
\ee
where we use the explicit form of the gamma matrices in appendix \ref{gammaappendix}.  Further, we have used the following ordering of the spin indices
\be
{+\dot{+}},\;\;{+\dot{-}},\;\;{+\dot{1}},\;\;{+ \dot{2}},\;\;
{-\dot{+}},\;\;{-\dot{-}},\;\;{-\dot{1}},\;\;{- \dot{2}},\;\;
{\dot{1} \dot{+}},\;\;{\dot{1} \dot{-}},\;\;{\dot{1}\dot{1}},\;\;{\dot{1} \dot{2}},\;\;
{\dot{2}\dot{+}},\;\;{\dot{2}\dot{-}},\;\;{\dot{2}\dot{1}},\;\;{\dot{2} \dot{2}}.
\ee
We use this ordering of the indices to write out gamma matrices explicitly so that any direct product structure is apparent.

The chirality matrices satisfy
\bea
&& \{\Gamma^5,\Gamma^{\mu}\}=0, \qquad \{\ti{\Gamma}^{\ti{5}},\ti{\Gamma}^{\ti{\mu}}\}=0, \\
&& [\Gamma^5,\ti{\Gamma}^{\ti{\mu}}]=0, \qquad [\ti{\Gamma}^{\ti{5}},\Gamma^{\mu}]=0,\nn \\
&& (\Gamma^5)^2=1, \qquad (\ti{\Gamma}^{\ti{5}})^2=1.\nn
\eea

\subsection{Choices of phases, and charge conjugation}
\label{thetaphases}

At this point, we would like to take advantage of the phases that dress each block of spin fields, $\theta_{pp},\theta_{p1},\theta_{1p}$ and $\theta_{11}$; see appendix \ref{app_theta_spinfields}.  It is tempting (and possible) to choose these phases so that the gamma matrices associated with left moving fermions appear as $\Gamma^\mu=\gamma^\mu\otimes 1$ and that the right moving counterparts appear as $\ti{\Gamma}^{\ti{\mu}}=\gamma^5\otimes\ti{\gamma}^{\ti{\mu}}$, yielding a direct product structure for the gamma matrices.  It would further be tempting (and possible) to assign the charge conjugation matrix appearing on the right hand side of $SS$ OPEs to have a direct product structure $C=c\otimes\ti{c}$, where $c$ and $\ti{c}$ are charge conjugation matrices for each set of gamma matrices.

One can enforce either of the above structures, but not both simultaneously.  Thus, we cannot bring the charge conjugation matrix and the gamma matrices individually into direct product form.  Because the charge conjugation matrix is used for raising and lowering spin indices we choose to simplify this structure rather than simplifying the structure of the gamma matrices.

A natural choice for the charge conjugation matrix is
\be
(S^{W\ti{X}})^{\dagger}=S_{W\ti{X}}=C_{W\ti{X} Y\ti{Z}}S^{Y\ti{Z}}= -\epsilon_{WY}\epsilon_{\ti{X}\ti{Z}}S^{Y\ti{Z}},
\ee
where the epsilons are $0$ when indices from different $SU(2)$s are involved.  This exactly mimics the reality condition for the fermions of the form $\psi^\dagger=-\epsilon\epsilon\psi$.  Thus, in all that follows, all charge indices are raised and lowered with $-i\epsilon=\sigma^2$.  This choice of charge conjugation matrix is accomplished by the following choice of phases
\bea
&& \theta_{pp}= \frac{\pi}{2}+\frac{\pi}{8}\begin{pmatrix}
+M_{55} & -M_{65} &-3M_{\ti{5} 5} & +3M_{\ti{6}5} \\
+3M_{56} & +M_{66} & +3M_{\ti{5}6} & -3M_{\ti{6}6} \\
+M_{5\ti{5}} & -M_{6\ti{5}} & +M_{\ti{5}\ti{5}} & -M_{\ti{6}\ti{5}} \\
-M_{5\ti{6}} & +M_{6\ti{6}} & +3M_{\ti{5}\ti{6}} & +M_{\ti{6}\ti{6}}
\end{pmatrix}\label{thetapp}, \\
&& \theta_{11}= -\frac{\pi}{2}+\frac{\pi}{8}\begin{pmatrix}
-3M_{55}     & +M_{65}      & -3M_{\ti{5} 5}     & -3M_{\ti{6}5} \\
-3M_{56}     & +M_{66}      & -3M_{\ti{5}6}      & -3M_{\ti{6}6} \\
+M_{5\ti{5}} & +M_{6\ti{5}} & -3M_{\ti{5}\ti{5}} & +M_{\ti{6}\ti{5}} \\
+M_{5\ti{6}} & +M_{6\ti{6}} & -3M_{\ti{5}\ti{6}} & +M_{\ti{6}\ti{6}}
\end{pmatrix}\label{theta11}, \\
&& \theta_{p1}= \frac{\pi}{2}+\frac{\pi}{8}\begin{pmatrix}
+M_{55}      & -M_{65}      & +3M_{\ti{5} 5}     & +3M_{\ti{6}5} \\
+3M_{56}     & +M_{66}      & -3M_{\ti{5}6}      & -3M_{\ti{6}6} \\
-M_{5\ti{5}} & +M_{6\ti{5}} & -3M_{\ti{5}\ti{5}} & +M_{\ti{6}\ti{5}} \\
-M_{5\ti{6}} & +M_{6\ti{6}} & -3M_{\ti{5}\ti{6}} & +M_{\ti{6}\ti{6}}
\end{pmatrix}\label{thetap1}, \\
&& \theta_{1p}= -\frac{\pi}{2}+\frac{\pi}{8}\begin{pmatrix}
-3M_{55}      & +M_{65}      & -M_{\ti{5} 5}     & +M_{\ti{6}5} \\
-3M_{56}     & +M_{66}      & -M_{\ti{5}6}      & +M_{\ti{6}6} \\
+3M_{5\ti{5}} & +3M_{6\ti{5}} & +M_{\ti{5}\ti{5}} & -M_{\ti{6}\ti{5}} \\
-3M_{5\ti{6}} & -3M_{6\ti{6}} & +3M_{\ti{5}\ti{6}} & +M_{\ti{6}\ti{6}}
\end{pmatrix}.\label{theta1p}
\eea
For this choice, the charge conjugation matrix then satisfies the following simple relations
\bea
&& C=C^T=C^{\dagger}=C^{-1}, \\
&& [C,\Gamma^{5}]=0,\qquad [C,\ti{\Gamma}^{\ti{5}}]=0.\nn
\eea
In addition to simplifying the charge conjugation matrix, this choice also makes the gamma matrices depend only of the antisymmetric part of the cocycle matrix $A_{ij}=M_{ij}-M_{ji}$, which the modulo arithmetic relations constrain. We display the spin fields with the above choice of phases in appendix \ref{app_theta_spinfields}.

Although the charge conjugation matrix is simple in form, the action on the gamma matrices is somewhat complicated because of the more complicated form of the gamma matrices.  The action of the charge conjugation matrix can be written as
\bea
&&(\Gamma^{\mu})^T=-\exp{\left[\frac{i\pi}{2}\left((-A_{5\ti{5}}-A_{56})\Gamma^5-A_{5\ti{6}}\Gamma^5\ti{\Gamma}^{\ti{5}}\right)\right]} C\Gamma^{\mu}C, \\
&& (\ti{\Gamma}^{\ti{\mu}})^T=-\exp{\left[\frac{i\pi}{2}\left((-A_{5\ti{5}}-A_{\ti{5}\ti{6}})\ti{\Gamma}^{\ti{5}} -A_{6\ti{5}}\Gamma^5\ti{\Gamma}^{\ti{5}}\right)\right]} C\ti{\Gamma}^{\ti{\mu}}C.\nn
\eea
This transformation, along with the extra phases, can easily be checked to leave the algebra (\ref{cliffalg}) invariant.  Further, it transforms the $SO(4)$ generators in the correct way
\be
C[\Gamma^{\mu},\Gamma^{\nu}]C=-\left([\Gamma^{\mu},\Gamma^{\nu}]\right)^T,\qquad C[\ti{\Gamma}^{\ti{\mu}},\ti{\Gamma}^{\ti{\nu}}]C=-\left([\ti{\Gamma}^{\ti{\mu}},\ti{\Gamma}^{\ti{\nu}}]\right)^T.
\ee

One may also work with multi fermions on the spin fields, and use gamma matrices to efficiently write
\bea
\psi^{\alpha \dot{A}}(t)S^{W\ti{X}}(0,0)\sim \frac{1}{t^{\frac{1}{2}}}\left(\Gamma^{\alpha \dot{A}}\right)^{W\ti{X}}\,_{Y\ti{Z}} S^{Y\ti{Z}}(0,0)+\cdots,\\
(\psi^{\alpha \dot{A}}\psi^{\beta \dot{B}})(t)S^{W\ti{X}}(0,0)\sim -\frac{1}{t}\left(\Gamma^{\left[ \alpha \dot{A}\right.}\Gamma^{\left.\beta \dot{B}\right]}\right)^{W\ti{X}}\,_{Y\ti{Z}} S^{Y\ti{Z}}(0,0)+\cdots,\nn\\
(\psi^{\alpha \dot{A}}\psi^{\beta \dot{B}}\psi^{\gamma \dot{C}})(t)S^{W\ti{X}}(0,0)\sim -\frac{1}{t^{\frac{3}{2}}}\left(\Gamma^{\left[ \alpha \dot{A}\right.}\Gamma^{\beta \dot{B}}\Gamma^{\left.\gamma \dot{C}\right]}\right)^{W\ti{X}}\,_{Y\ti{Z}} S^{Y\ti{Z}}(0,0)+\cdots,\nn\\
(\psi^{\alpha \dot{A}}\psi^{\beta \dot{B}}\psi^{\gamma \dot{C}}\psi^{\delta \dot{D}})(t)S^{W\ti{X}}(0,0)\sim \frac{1}{t^{2}}\left(\Gamma^{\left[ \alpha \dot{A}\right.}\Gamma^{\beta \dot{B}}\Gamma^{\gamma \dot{C}}\Gamma^{\left.\delta \dot{D}\right]}\right)^{W\ti{X}}\,_{Y\ti{Z}} S^{Y\ti{Z}}(0,0)+\cdots,\nn
\eea
where the antisymmetrization is done on pairs of indices with weight one.  For example
\bea
(\Gamma^{\left[ \alpha \dot{A}\right.}\Gamma^{\beta \dot{B}}\Gamma^{\left.\gamma \dot{C}\right]})&=&\frac{1}{3!} \Bigg(\Gamma^{\alpha \dot{A}}\Gamma^{\beta \dot{B}}\Gamma^{\gamma \dot{C}}+\Gamma^{\beta \dot{B}}\Gamma^{\gamma \dot{C}}\Gamma^{\alpha \dot{A}}+\Gamma^{\gamma \dot{C}}\Gamma^{\alpha \dot{A}}\Gamma^{\beta \dot{B}} \nn \\
&& \qquad -\Gamma^{\beta \dot{B}}\Gamma^{\alpha \dot{A}}\Gamma^{\gamma \dot{C}}-\Gamma^{\alpha \dot{A}}\Gamma^{\gamma \dot{C}}\Gamma^{\beta \dot{B}}-\Gamma^{\gamma \dot{C}}\Gamma^{\beta \dot{B}}\Gamma^{\alpha \dot{A}}\Bigg).
\eea
The same relations hold for the antiholomorphic fermions with $\psi\rightarrow \ti{\psi}$, $t\rightarrow \bar{t}$, $\Gamma\rightarrow \ti{\Gamma}$, $\alpha \dot{A}\rightarrow\dot{\alpha}\dot{A}$, etc.  Further, the minus signs in front of the various terms can be accounted for by operating the $\psi$ operators in the reverse order than they appear, i.e. the fermion furthest to the right `operates on the spin field first'.  Any multifermions that involve fermions from both holomorphic and antiholomorphic sectors may be found as the product of the above term because the two sectors do not share OPEs.  For example,
\be
(\ti{\psi}^{\dot{\alpha}\dot{A}}\psi^{\beta\dot{B}}\psi^{\gamma\dot{C}}) (t,\bar{t})S^{W\ti{X}}(0,0)\sim -\frac{1}{t\bar{t}^{\frac{1}{2}}}\left(\ti{\Gamma}^{\dot{\alpha}\dot{A}} \Gamma^{\left[\beta\dot{B}\right.}\Gamma^{\left.\gamma\dot{C}\right]}\right)^{W\ti{X}}\,_{Y\ti{Z}} S^{Y\ti{Z}}(0,0)+\cdots.
\ee

\subsection{Four spin field correlator}

We now turn to the question of four-point functions, and in particular, we concentrate on the four spin field correlator:
\be
A_4(t_i,\tb_i)^{S\tilde{T}U\tilde{V}W\tilde{X}Y\tilde{Z}}=\langle S^{S\tilde{T}}_1S^{U\tilde{V}}_2S^{W\tilde{X}}_3S^{Y\tilde{Z}}_4\rangle,
\ee
where above we write subscripts $1,2,3,4$ to denote the location $(t_i,\tb_i)$ of the spin field insertion (note: we use the subscript $i$ on locations, not mistaking these for copy indices).

The amplitude $A_4$ should be invariant under $SU(2)$ transformations, even though it has free raised indices.  It is therefore natural to use the gamma matrices to express the answer.  We find it particularly convenient to use the hermitian gamma matrices in equations (\ref{gamma34})-(\ref{gammat1234}) simply because the indices $\mu, \tilde{\mu}$ are raised and lowered with $\delta^{\mu \nu}$ and $\delta^{\ti{\mu}\ti{\nu}}$ respectively, making it relatively straightforward to contract these indices.

Paying attention to holomorphic gamma matrices only, there is a natural set
\be
C^{S\tilde{T}U\tilde{V}}, \qquad (\Gamma^\mu C)^{S\tilde{T}U\tilde{V}}, \qquad (\Gamma^{[\mu \nu]}C)^{S\tilde{T}U\tilde{V}}, \qquad (\Gamma^{[\mu \nu \rho]}C)^{S\tilde{T}U\tilde{V}}, \qquad (\Gamma^{[\mu \nu \rho \sigma]}C)^{S\tilde{T}U\tilde{V}}.
\ee
where again, antisymmetrization is done weight one.  Note that the last two could be replaced using $\Gamma^{5}$ and and $\epsilon_{\mu \nu \rho \sigma}$ to reduce the number of free indices.  We recognize these as the usual projections of the product of two spin representations onto the scalar, vector, antisymmetric tensor, pseudo-vector, and pseudo-scalar representations of $SO(4)$.  One may be more careful, and decompose the antisymmetric tensor representation into self dual and anti self dual parts by considering $1/2(1\pm\Gamma^5)\Gamma^{\mu \nu}$, however, we find that this distinction is not needed in what follows: the self-dual and anti-self dual pieces always get the same coefficients.  There are the analogous combinations on the antiholomorphic side, and so we find a total of 25 gamma matrix structures that are important.

The total number of such gamma matrices, entering distinct $\mu,\ti{\mu}$,  is $(1+4+6+4+1)\times(1+4+6+4+1)=16^2$.  Due to the trace orthogonality, these provide a complete basis on the set of $16^2$ raised indices ${S\ti{T}U\ti{V}}$.

These gamma matrices provide the projections on a pair of doublet spin indices into the relevant $SO(4)\times SO(4)$ representation.  For example, the projection
\be
(C\Gamma^{\mu}\ti{\Gamma}^{\ti\mu\ti\nu\ti\rho})_{U\ti V S\ti T}S_1^{S\ti{T}}S_2^{U\ti{V}}\label{projectionexample}
\ee
gives the vector $\times$ pseudo-vector representation of $SO(4)\times SO(4)$.  We can make similar projections of the second set of spin fields appearing in the four point correlator as well.  However, now that we have each set projected onto irreducible representations\footnote{up to the self-dual/anti self dual comment before.  If we included this distinction, there would be 36 total structures to include.}, we can guess that these must come in the singlet combination, i.e. we expect the $S_1S_2$ projection to match the $S_3 S_4$ representation to make a singlet.  Matching only pseudovectors with other pseudovectors is assumed so that the correlator is parity invariant as well.  Hence, we take the ansatz
\bea
&& \kern-1em A_4(t_i,\tb_i)^{S\tilde{T}U\tilde{V}W\tilde{X}Y\tilde{Z}}=\nn \\
&& \frac{1}{16^2}\sum_{m=0,n=0}^{m=4,n=4}f_{mn} \frac{1}{m!}\frac{1}{n!}\left(\Gamma^{\mu_1\cdots \mu_m} \ti{\Gamma}^{\ti{\nu}_1 \cdots \ti{\nu}_n}C\right)^{S\tilde{T}U\tilde{V}}
\left(\Gamma_{\mu_1\cdots \mu_m} \ti{\Gamma}_{\ti{\nu}_1 \cdots \ti{\nu}_n}C\right)^{W\ti{X}Y\ti{Z}}, \label{4pointansatz}
\eea
where the functions $f_{mn}$ are functions of the $t_i,\tb_j$.  The labels $mn$ simply label the number of antisymmetrized gamma matrices that appear from the left vs. right sectors.  In the cases where the index goes to 0, this simply means to write no gamma matrices.  For example, the $m=1,n=0$ term in the sum is simply $f_{10} (\Gamma^{\mu}C)(\Gamma_{\mu}C)$ (note that here, and in other places of the text, we suppress the spin indices for brevity).

\begin{figure}[\ht!]
\begin{center}
\includegraphics[width=0.4\textwidth]{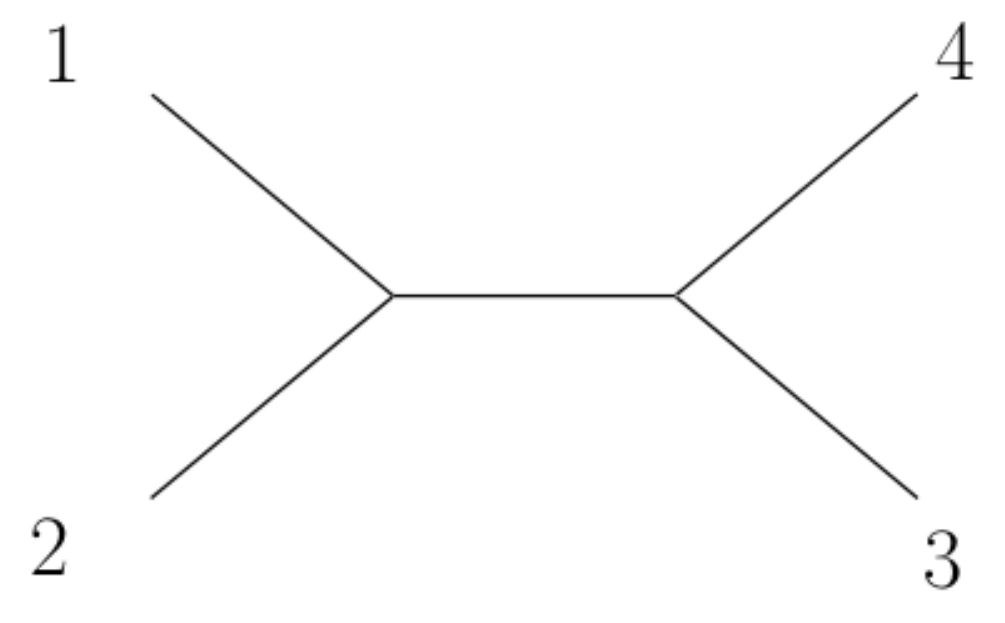}
\end{center}
\caption{The partial wave decomposition easiest to explore, given the form of the right hand side of (\ref{4pointansatz}).  Other partial wave decompositions are related to this one through Fierz identities.} \label{figpwave}
\end{figure}

Each of the above singlet gamma matrix structures can be shown to vanish unless the sum of momenta is zero, i.e. $k_1+k_2+k_3+k_4=0$, agreeing with expectations for the correlator.  Further, the above presentation is particularly convenient for the partial wave decomposition shown in figure \ref{figpwave}
which we use to detect mixing between operators in different twist sectors in the presence of a deformation operator \cite{Burrington:2012yq,Burrington:2012yn}.

We must now fix the 25 coefficient functions $f_{mn}$ appearing in (\ref{4pointansatz}).  These may be found by doing a subset of the projections similar to (\ref{projectionexample}).  However, in this projection, we do not have to worry about leaving the indices on the multi-gamma matrices arbitrary.  This is because the coefficient of, for example, the $(\Gamma^1\ti\Gamma^{\ti2}C)(\Gamma_1\ti\Gamma_{\ti{2}}C)$ type term should be the same as the $(\Gamma^2\ti\Gamma^{\ti4}C)(\Gamma_2\ti\Gamma_{\ti{4}}C)$ type term to combine these into a singlet.  One can show that this is actually the case, and so our ansatz is in fact consistent.  Hence, we only need to do 25 projections, using one gamma matrix from each representation.  Thus, for example, to find the function $f_{23}$ we take
\be
f_{23}=(C\Gamma^{12}\ti{\Gamma}^{\ti1}\ti{\Gamma}^{\ti 5})_{U\ti V S\ti T}(C\Gamma_{12}\ti{\Gamma}_{\ti 1}\ti{\Gamma}^{\ti 5})_{Y\ti Z W\ti X}A_4(t_i,\tb_i)^{S\tilde{T}U\tilde{V}W\tilde{X}Y\tilde{Z}}.
\ee
The individual terms in $A_4$ are computed by writing out the spin fields, moving all phases and cocycle operators to the left, and then using the na\"{\i}ve correlator
\be
\langle e^{ik_1\cdot \phi}_1e^{ik_2\cdot \phi}_2e^{ik_3\cdot \phi}_3e^{ik_4\cdot \phi}_4\rangle = \delta_{\sum k_i=0} \prod_{i<j} (t_{ij})^{k_{i,L}\cdot k_{j,L}}(\tb_{ij})^{k_{i,R}\cdot k_{j,R}},\label{naive}
\ee
where as usual, $k_L$ refers to the momentum in front of the holomorphic $\phi$ operators, and $k_R$ refers to the momentum in front of the antiholomorphic $\ti{\phi}$ operators.  We find it convenient to express these projections in terms of the functions
\be
f_{\pm}=(\sqrt{t_{13}}\sqrt{t_{24}}\pm\sqrt{t_{14}}\sqrt{t_{23}}), \qquad \bar{f}_{\pm}=(\sqrt{\tb_{13}}\sqrt{\tb_{24}}\pm\sqrt{\tb_{14}}\sqrt{\tb_{23}}).
\ee
Further, we find that the functions obey $(f_{ij})^*=f_{ji}$ (star acts as $t\leftrightarrow \tb$), and so we will only list the 15 independent functions:
\bea
&& f_{00}= \frac{16}{|F|}\left|f_+\right|^4 \nn \\
&& f_{10}= \frac{16}{|F|}\sqrt{t_{12}}\sqrt{t_{34}}f_+\bar{f}_+^2 \nn \\
&& f_{20}= -\frac{16}{|F|}f_+f_-\bar{f}_+^2\qquad f_{11}=-\frac{16}{|F|}||t_{12}||t_{34}|\left|f_+\right|^2 \nn \\
&& f_{30}= -\frac{16}{|F|}\sqrt{t_{12}}\sqrt{t_{34}}f_+^2\bar{f}_-
\qquad f_{21}= -\frac{16}{|F|}f_+f_- \sqrt{\tb_{12}}\sqrt{\tb_{34}}\bar{f}_+ \nn \\
&& f_{40}=\frac{16}{|F|}f_{-}^2\bar{f}_+^2 \qquad f_{31}=\frac{16}{|F|}|t_{12}||t_{34}|f_-\bar{f}_+ \qquad
f_{22}= \frac{16}{|F|}|f_+f_-|^2\\
&& f_{41}= \frac{16}{|F|} f_-^2\sqrt{\tb_{12}}\sqrt{\tb_{34}}\bar{f}_+\qquad f_{32}=\frac{16}{|F|}\sqrt{t_{12}}\sqrt{t_{34}}f_-\bar{f}_+\bar{f_-} \nn \\
&& f_{42}=-\frac{16}{|F|}f_-^2\bar{f}_+\bar{f}_- \qquad f_{33}=-\frac{16}{|F|}|t_{12}||t_{34}||f_-|^2 \nn \\
&& f_{43}=-\frac{16}{|F|}f_-^2\sqrt{\tb_{12}}\sqrt{\tb_{34}}\bar{f}_-\nn \\
&& f_{44}=\frac{16}{|F|}|f_-|^4 \nn
\eea
where
\be
|F|=|t_{12}||t_{13}||t_{14}||t_{23}||t_{24}||t_{34}|.
\ee
Given these functions, one can indeed verify that (\ref{4pointansatz}) gives the four point function exactly.\footnote{This was done by verifying all 1,296 separate entries using Maple.}  Thus, we have verified that the four point function does not transform under the $SO(4)\times SO(4)$ symmetry; it is a singlet.

\section{Consistency of correlators in the orbifold theory}
\subsection{Crossing symmetry, an integral lattice on the cover.}

It is well known that OPEs that contain branch cuts, such as those shown above, contain ambiguities when requiring crossing symmetry (i.e. the conformal bootstrap).  In string theory, these branch cuts are removed by combinations of physical state conditions and the GSO projection, which restrict the theory to an integral lattice.  Here, we simply mention an obvious integral lattice.  If we take the operation $\Gamma=\Gamma^5\ti{\Gamma}^{\ti 5}$, we can restrict ourselves to the $\Gamma=1$ sector of the theory.  This restricts the set of spin fields to $S^{+/-,\dot{+}/\dot{-}}$ and $S^{\dot{1}/\dot{2},\dot{1}/\dot{2}}$ as well as restricting to even number fermion excitations (only bi-fermion, quad-fermion, etc. operators are allowed).  This makes a self consistent set of vertex operators with crossing symmetry.  Note that the momentum vectors associated with the $+/-$ sectors (with or without dots) are perpendicular to those in the $\dot{1}/\dot{2}$ sector.  Because of this, these operators share no singular OPEs.  In fact, it is natural for these to commute.  This is easy to check using the 4 point functions above:
\bea
&& \langle S^{+\dot{+}}_1 S^{-\dot{-}}_2 S^{\dot{1}\dot{1}}_3 S^{\dot{2}\dot{2}}_4\rangle = \frac{1}{|t_{12}||t_{34}|}, \\
&& \langle S^{+\dot{+}}_1 S^{\dot{1}\dot{1}}_3 S^{-\dot{-}}_2 S^{\dot{2}\dot{2}}_4\rangle =\frac{\exp\left(-\frac{i\pi}{2} \left( A_{56}+A_{5\ti{6}}-A_{6\ti{5}}+A_{\ti 5 \ti 6}\right)\right)}{|t_{12}||t_{34}|},
\eea
and so we get the additional constraint that
\be
A_{56}+A_{5\ti{6}}-A_{6\ti{5}}+A_{\ti 5 \ti 6}\equiv 0 \; \Mod{4},
\ee
which is clearly consistent with all earlier modulo arithmetic constraints (\ref{amod2const}), (\ref{ppsector})-(\ref{11sector}): neither $A_{56}$ nor $A_{\ti 5 \ti 6}$ appeared in any constraints modulo 4 until now.  Thus, given $A_{5\ti{6}}$ and $A_{6\ti{5}}$, the above constraint can be suitably satisfied by some choice of $A_{56}$ and $A_{\ti 5 \ti 6}$.  We choose to have these operators commute to agree with spin statistics; $h-\tilde{h}=0$ for all operators involved.

One may worry that the above constraint on the lattice is too severe, and that some states of interest will be excluded.  If this is true, one may hope that there are other properties of the orbifold that give more structure to the correlators that one would actually compute.  We will see in the next subsection that this is actually the case.

Nevertheless, we feel it is important to note that for simple operators, such as those considered in our earlier work \cite{Burrington:2012yq}, the above integral lattice will be sufficient to deal with all operators of interest.  For example, the lift to the covering surface of the (anti) chiral primary twist two operators $\sigma^{\pm,\dot{\pm}}$ all appear in this class, as well as the deformation operator's lift to the cover, as well as the spinless singlet bosonic operator considered in 
\cite{Burrington:2012yq}.

\subsection{Beyond the integral lattice: simple three point functions}\label{3pf_sigma2}

Here we comment on extra structure that the orbifold theory has beyond a generic CFT with spin fields.  First, let us consider a correlator that obviously has a branch cut,
\be
A_{3,t}=\langle \psi^{-\dot{1}}(t_0)\, S^{+\dot{+}}(t_1,\tb_1) \, S^{\dot{2} \dot{-}}(t_2,\tb_2)\rangle. \label{branchexample}
\ee
The field $\psi^{-\dot{1}}$ has a branch cut with respect to the two spin fields.  Let us try to imagine where this might come from in the base space.  An operator of the form
\be
A_{3,z}=\langle (\psi_1^{-\dot{1}}+\psi_2^{-\dot{1}})(z_0)\, \sigma^{+\dot{+}}_{12}(z_1,\zb_1)\, \sigma^{\dot{2} \dot{-}}_{12}(z_2,\zb_2)\rangle \label{baseexample}
\ee
will lead to computations like (\ref{branchexample}) in the cover.
Further, (\ref{baseexample}) is clearly well defined in the base.  As the fermion operator circles either twist field the two copy labels interchange ${1}\leftrightarrow {2}$, leaving the operator invariant.  Hence, even though the covering version appears to have branch cuts, the base space clearly does not.  In fact, one can trace this back to the conformal weight of the fermion, and we can see that {bare calculations of the form (\ref{branchexample}) do not appear in the cover}, given a well defined correlator on the base.

To explore this further, we consider the map from the base to the cover \cite{Burrington:2012yn,Burrington:2012yq,Lunin:2001pw,Lunin:2000yv}.  The most general map that takes $z_1\rightarrow t_1$ and $z_2\rightarrow t_2$ with the proper ramification is
\be
\frac{(z-z_1)}{(z-z_2)}=\left(a\frac{(t-t_1)}{(t-t_2)}\right)^2.
\ee
There are two images of the point $z_0$, and these are given by
\bea
t_+=\frac{t_1z_{02}a^2-t_2z_{01}+at_{12}\sqrt{z_{02}}\sqrt{z_{01}}} {z_{02}a^2-z_{01}},\\
t_-=\frac{t_1z_{02}a^2-t_2z_{01}-at_{12}\sqrt{z_{02}}\sqrt{z_{01}}} {z_{02}a^2-z_{01}}.
\eea
Note that we have written all square roots with $z_{ij}$ such that $i<j$, obeying a `radially normal ordered' rule, just the same as the  correlator for multiple exponentials (\ref{naive}).

To lift the computation (\ref{baseexample}) to a calculation resembling (\ref{branchexample}), we follow the rules in \cite{Lunin:2001pw,Lunin:2000yv,Burrington:2012yn}. For the normalization of the lift operators, we need the expansions near the ramified points on the cover.  In general, these will be of the form
\be
z-z_i=b(t-t_i)^n+\cdots.
\ee
We see that this is the map $z-z_i=(\hat{t}-\hat{t}_i)^n$ followed by a conformal scaling $(\hat{t}-\hat{t}_i)=b^{1/n}(t-t_i)$.  The normalized twist operators are defined by the first map, but the conformal scaling introduces a factor of $b^{h/n}\bar{b}^{\tilde{h}/n}$ if there is an operator at that point of weight $(h,\tilde{h})$.  For more details, see \cite{Lunin:2001pw}. We have spin fields at the ramifield point in the $t$ plane, and so the lift to the cover gets dressed by terms of this nature.

Following these rules, we lift
\bea
&& \langle (\psi_1^{-\dot{1}}+\psi_2^{-\dot{1}})(z_0)\, \sigma^{+\dot{+}}_{12}(z_1,\zb_1)\, \sigma^{\dot{2} \dot{-}}_{12}(z_2,\zb_2)\rangle\\
&& \qquad \rightarrow \Biggl\langle \left( \left. \frac{\pa z}{\pa t} \right| _{t_+}\right)^{-1/2} \psi^{-\dot{1}}(t_+)\, b_1^{-1/8}\bar{b}_1^{-1/8}S^{+\dot{+}}(t_1,\tb_1) \, b_2^{-1/8}\bar{b}_2^{-1/8} S^{\dot{2} \dot{-}}(t_2,\tb_2)\Biggr\rangle \label{lifteq}\nn\\
&& \qquad\;+\,\Biggl\langle \left( \left. \frac{\pa z}{\pa t} \right| _{t_-}\right)^{-1/2} \psi^{-\dot{1}}(t_-)\, b_1^{-1/8}\bar{b}_1^{-1/8} S^{+\dot{+}}(t_1,\tb_1) \, b_2^{-1/8}\bar{b}_2^{-1/8} S^{\dot{2} \dot{-}}(t_2,\tb_2)\Biggr\rangle,\nn
\eea
where $\rightarrow$ indicates a lift to the cover, neglecting the term from the conformal anomaly \cite{Lunin:2001pw,Lunin:2000yv} (which we will add later).

The partial derivatives $\frac{\pa{z}}{\pa t}$ terms can be written succinctly as
\be
\left( \left. \frac{\pa z}{\pa t} \right| _{t_+}\right)^{-1}=\frac{1}{2}\frac{t_{+1}t_{+2}}{t_{12}}\frac{z_{12}} {z_{01}{z_{02}}}, \qquad \left( \left. \frac{\pa z}{\pa t} \right| _{t_-}\right)^{-1}=\frac{1}{2}\frac{t_{-1}t_{-2}}{t_{12}}\frac{z_{12}} {z_{01}{z_{02}}},
\ee
where $t_{\pm 1}=t_{\pm}-t_1$.  The form of this is highly suggestive.  Note that all terms are written using normal ordering that is implied by (\ref{lifteq}).  In fact, this is the rule that we shall use: the radial normal ordering implemented on the base shall be lifted to the cover, even if this is not `radial' in the cover.  This will be more important in our next calculation.

Paying attention to the first term on the right hand side of (\ref{lifteq})
\bea
&&\Biggl\langle \left( \left. \frac{\pa z}{\pa t} \right| _{t_+}\right)^{-1/2} \psi^{-\dot{1}}(t_+)\, b_1^{-1/8}\bar{b}_1^{-1/8}S^{+\dot{+}}(t_1,\tb_1) \,b_2^{-1/8}\bar{b}_2^{-1/8} S^{\dot{2} \dot{-}}(t_2,\tb_2)\Biggr\rangle\\
&& \qquad \qquad=\frac{1}{\sqrt{2}}\frac{\sqrt{t_{+1}}\sqrt{t_{+2}}}{\sqrt{t_{12}}} \frac{\sqrt{z_{12}}}{\sqrt{z_{01}}\sqrt{z_{02}}}\Biggl\langle \psi^{-\dot{1}}(t_+)\, |b_1|^{-1/4}S^{+\dot{+}}(t_1,\tb_1) \, |b_2|^{-1/4}S^{\dot{2} \dot{-}}(t_2,\tb_2)\Biggr\rangle\nn\\
&& \qquad \qquad = \frac{1}{\sqrt{2}}\frac{\sqrt{t_{+1}}\sqrt{t_{+2}}}{\sqrt{t_{12}}} \frac{\sqrt{z_{12}}}{\sqrt{z_{01}}\sqrt{z_{02}}} |b_1|^{-1/4}|b_2|^{-1/4}\left(\frac{e^{i\delta}}{\sqrt{t_{+1}}\sqrt{t_{+2}}\sqrt{\tb_{23}}}\right)\nn\\
&& \qquad \qquad = \frac{1}{\sqrt{2}}\frac{|b_1|^{-1/4}|b_2|^{-1/4}}{|{t_{12}}|} \frac{\sqrt{z_{12}}}{\sqrt{z_{01}}\sqrt{z_{02}}} e^{i\delta},\nn
\eea
where $e^{i\delta}$ is a calculable phase (however, we shall not need it here).  We will discuss how to align branch cuts in our next calculation, however, we note here that our prescription will be to have the $\sqrt{t_{+1}}$ owing to the conformal weight transformation of $\psi$ align with the branch cut in $\sqrt{t_{+1}}$ coming from the correlator, and likewise for $\sqrt{t_{+2}}$ and $\sqrt{t_{12}}$, leading to the third equality above.  Our result here will not depend on whether we cancel these terms, or include a possible $-1$ for choosing different branches.  Note also that there will be a similar expression for the second term on the right hand side of (\ref{lifteq}), with $t_+\rightarrow t_-$.

Next, we have to add the results from the two images at $t_+$ and $t_-$.  Therefore, we must make sense of the square roots in front of each of these terms to be able to add them.  However, there is a simple way to relate them using the base space.  Recall that while circling one twist field, $\psi_1\rightarrow \psi_2$.  So, the calculation at $t_-$ is related to the one at $t_+$ by taking $z_{01}\rightarrow e^{i\theta}z_{01}$ and continuously varying $\theta$ from $0$ to $2\pi$.  The other expressions vary as well, but since the point $z_0$ is only circling $z_1$, none of the other terms cross to a different branch.  Doing so takes $\sqrt{z_{01}}\rightarrow-\sqrt{z_{01}}$, and maps $t_+\rightarrow t_-$.  However, the two terms do not have any dependence left on $t_+$ or $t_-$, and so the two expressions cancel.  Hence, we find
\be
\langle (\psi_1^{-\dot{1}}+\psi_2^{-\dot{1}})(z_0)\, \sigma^{+\dot{+}}_{12}(z_1,\zb_1)\, \sigma^{\dot{2} \dot{-}}_{12}(z_2,\zb_2)\rangle
\rightarrow 0
\ee
Thus, the structure from the base space avoids the branch cuts in a calculation of the type (\ref{branchexample}) in the most simple way: it sets it to zero.


One may be curious about using the operator with signs that counteract this cancellation, i.e. $\psi_1-\psi_2$.  The additional sign would mean that the lift would give a non-zero answer.  However, such an operator is not single valued with respect to the twist operation, and so will not come up in calculations for the orbifold CFT.  Operators of the type $\psi_1-\psi_2$ are only well defined inside an integral, evaluated on a contour surrounding a twist 2 operator, with a half integer power of $z$ multiplying them.  These fields are used to excite a twist, and are not operators corresponding to states in their own right.  They must always appear as a mode acting on a twist.

Before we move on, we want to justify our choice of branches for various factors of $\sqrt{t_{+1}}$ in the above calculation.  We claim that this aligning of branch cuts is the last piece necessary in lifting to the cover, and is in fact nothing new.  Usually, when defining the NS sector on the cylinder, one must use half integer Fourier modes to pick off the modes of a fermion $\psi$.  Thus, the single valued combination $\sqrt{e^{-iw}}\psi(w)$ occur when defining the modes of $\psi$ on the cylinder (we use $w$ as the coordinate on the cylinder).   When mapping to the plane using the usual exponential map, there is an introduction of a $\sqrt{z}$ owing to the conformal weight of $\psi$.  This is taken to have the same branch cut as the $\sqrt{e^{-iw}}=\sqrt{z}$ that appears in the Fourier transform, so that the modes on the plane, using the residue theorem, and on the cylinder, using the Fourier transform, are in fact the same.  In our case, aligning the branch cuts of $t_{+1}$ coming from the conformal weight and the correlator is an identical statement, and in part defines what we mean by the lift to the cover.

However, in the above example of the usual story for the NS sector, the map is one to one.  In our case the map is one to many copies.  We will choose the following prescription.  We choose one image of a base point to define the first patch in the cover \cite{Burrington:2012yn}.  When we lift to the cover, there may be certain branch cuts that are introduced owing to the conformal weight of the operator if the operator is fermionic.  These will be aligned with all branch cuts coming from interactions with all spin fields present.  This defines any operator in patch 1.  Operators in other patches are constructed by defining them in patch 1, and then transporting them to the appropriate point in the other patch.  This transportation from one image to the next is exactly a circling of the appropriate twist field.  The calculation at the new point may have picked up phases or signs owing to some $\sqrt{z_{ij}}$ terms.

We claim that this prescription is independent of the patch labeled as the first patch in the cover.  Above, it would not have mattered if we had picked patch 1 as being associated with $t_-$.  There still would have been a relative -1 between calculations setting it to zero.  In the general twist $n$ case, we would expect $n$ copies, but each calculation would be off by $e^{2\pi i/n}$, and the sum over such terms would be again zero.  On the other hand, if we choose an operator from patch one times an operator from patch two, there will be an overall phase that is prescribed in this manner.

We now turn to a calculation that gives a non-zero answer and also elucidates the above comments.  We consider
\be
A_{3,z}'=\langle (\psi_1^{\alpha \dot{A}}\psi_2^{\beta \dot{B}})(z_0)\, \sigma^{W\ti{X}}_{12}(z_1,\zb_1)\, \sigma^{Y\ti{Z}}_{12}(z_2,\zb_2)\rangle
\ee
We note that as this composite operator circles $\sigma_{12}$, the two subscripts $1$ and $2$ are interchanged.  To move them back, we permute the fermions past one another, giving a minus sign, resulting in the same correlator, with a minus sign, and the superscripts $\alpha \dot{A}$ and $\beta \dot{B}$ interchanged.  Thus, to have a well defined operator, the superscripts must be antisymmetric under interchange.  One can achieve this by taking the antisymmetric combination of $\dot{A}$ and $\dot{B}$, and the symmetric combination of $\alpha$ and $\beta$.  This results in an $SU(2)_L$ triplet, and an $SU(2)_2$ singlet.  One could, of course, have symmetrized these the other way (i.e. using an $\epsilon_{\alpha \beta}$).  Thus, we find a sensible computation is
\bea
A_{3,z}'&=&\Bigl\langle \epsilon_{\dot{A}\dot{B}}\left(\psi_1^{+ \dot{A}}\psi_2^{+ \dot{B}}\right)(z_0)\, \sigma^{-\dot{+}}_{12}(z_1,\zb_1)\, \sigma^{- \dot{-}}_{12}(z_2,\zb_2)\Bigr\rangle\\
&=&\Bigl\langle \left(\psi_1^{+ \dot{1}}\psi_2^{+ \dot{2}}+\psi_2^{+ \dot{1}}\psi_1^{+ \dot{2}}\right)(z_0)\, \sigma^{-\dot{+}}_{12}(z_1,\zb_1)\, \sigma^{- \dot{-}}_{12}(z_2,\zb_2)\Bigr\rangle.\nn
\eea
This is clearly invariant when circling the twist operator.  We lift the computation to the covering surface, finding
\bea
&&\kern-3.2em \Bigl\langle\epsilon_{\dot{A}\dot{B}} \left(\psi_1^{+ \dot{A}}\psi_2^{+ \dot{B}}\right)(z_0)\, \sigma^{-\dot{+}}_{12}(z_1,\zb_1)\, \sigma^{- \dot{-}}_{12}(z_2,\zb_2)\Bigr\rangle \rightarrow \\
\qquad   && \kern-3.2em \Bigl\langle\epsilon_{\dot{A}\dot{B}}\left( \left. \frac{\pa z}{\pa t} \right| _{t_+}\right)^{-1/2}\psi^{+\dot{A}}(t_+)\left( \left. \frac{\pa z}{\pa t} \right| _{t_-}\right)^{-1/2}\psi^{+\dot{B}}(t_-) |b_1|^{-1/4}S^{-\dot{+}}(t_1,\tb_1)|b_2|^{-1/4}S^{-\dot{-}}(t_2,\tb_2)\Bigr\rangle.\nn
\eea
We again have different square roots at different points.  We now make use of the comments above.  To define the operator $\left( \left. \frac{\pa z}{\pa t} \right| _{t_-}\right)^{-1/2}\psi^{+\dot{B}}(t_-)$, which is in the second patch, we start with it in the first patch $\left( \left. \frac{\pa z}{\pa t} \right| _{t_+}\right)^{-1/2}\psi^{+\dot{B}}(t_+)$ and continuously deform it to the point $t_1$ by circling one of the spin fields, for concreteness, we pick the one at $z_1$.  Owing to the square root of $z_{01}$ appearing in the conformal weight transformation, the operator picks up a minus sign.  However, the branch cuts between $\psi^{+\dot{B}}(t)$ and the spin fields still cancel, since both are deformed together.  This corresponds to a $\sqrt{t_{+1}}$ from the transformation term canceling a $1/\sqrt{t_{+1}}$ in the first patch deforming to a $\sqrt{t_{-1}}$ cancelling a $1/\sqrt{t_{-1}}$ in the second patch.  Thus, there is a sign introduced in the lift.  As promised, it would not matter whether we used $t_{-}$ to define the first patch: a relative minus sign would have been introduced when defining the operator at location $t_{+}$ by continuously deforming one starting from $t_-$.  Thus, we find
\bea
&&\kern-1em\Bigl\langle\epsilon_{\dot{A}\dot{B}}\left( \left. \frac{\pa z}{\pa t} \right| _{t_+}\right)^{-1/2}\psi^{+\dot{A}}(t_+)\left( \left. \frac{\pa z}{\pa t} \right| _{t_-}\right)^{-1/2}\psi^{+\dot{B}}(t_-) |b_1|^{-1/4}S^{-\dot{+}}(t_1,\tb_1)|b_2|^{-1/4}S^{-\dot{-}}(t_2,\tb_2)\Bigr\rangle \nn \\
&&=-\frac{1}{\sqrt{2}}\frac{\sqrt{t_{+1}}\sqrt{t_{+2}}}{\sqrt{t_{12}}} \frac{\sqrt{z_{12}}}{\sqrt{z_{01}}\sqrt{z_{02}}}\frac{1}{\sqrt{2}} \frac{\sqrt{t_{-1}}\sqrt{t_{-2}}}{\sqrt{t_{12}}} \frac{\sqrt{z_{12}}}{\sqrt{z_{01}}\sqrt{z_{02}}}|b_1|^{-1/4}|b_2|^{-1/4} \nn \\
&&\qquad \qquad\qquad \qquad \qquad\times \bigl\langle \epsilon_{\dot{A}\dot{B}}\psi^{+\dot{A}}(t_+)\psi^{+\dot{B}}(t_-) S^{-\dot{+}}(t_1,\tb_1)S^{-\dot{-}}(t_2,\tb_2)\bigr\rangle\nn\\
&&=\frac{z_{12}}{z_{01}z_{02}|t_{12}|}|b_1|^{-1/4}|b_2|^{-1/4}.
\eea
From our transformation, it is easy to read off the expansions of $b_1$ and $b_2$.  We expand near $t_1$ and $t_2$ to find
\be
z-z_1=a^2\frac{z_{12}}{(t_{12})^2}(t-t_1)^2+\cdots,\qquad z-z_2=-\frac{1}{a^2}\frac{ z_{12}}{(t_{12})^2}(t-t_2)^2+\cdots.
\ee
Thus, $|b_1|=|a|^2|b|$ and $|b_2|=\frac{1}{|a|^2}|b|$ where $b=\frac{z_{12}}{(t_{12})^2}$.  Plugging this in, we find
\be
\langle \epsilon_{\dot{A}\dot{B}} (\psi_1^{+ \dot{A}}\psi_2^{+ \dot{B}})(z_0)\, \sigma^{-\dot{+}}_{12}(z_1,\zb_1)\, \sigma^{-\dot{-}}_{12}(z_2,\zb_2)\rangle\rightarrow \frac{z_{12}}{z_{01} z_{02}|z_{12}|^{1/2}}.
\ee
To write a strict equality, we must include a factor coming from the conformal anomaly.  This was calculated in \cite{Lunin:2000yv} for our map in the case $a=1,z_1=0,t_1=0,t_2=1$.  We take that our twist fields are normalized, and so this term results in an additional factor of $|z_{12}|^{-\frac{c}{6}\left(n-1/n\right)}$ where for us $c=6$ and $n=2$, giving the final result
\be
\langle \epsilon_{\dot{A}\dot{B}} (\psi_1^{+ \dot{A}}\psi_2^{+ \dot{B}})(z_0)\, \sigma^{-\dot{+}}_{12}(z_1,\zb_1)\, \sigma^{-\dot{-}}_{12}(z_2,\zb_2)\rangle=\frac{1}{z_{01} z_{02}  \zb_{12}}.
\ee
Note that this has the proper form for a three point function between (quasi) primary operators of conformal weights $(1,0)$, $(1/2,1/2)$, and $(1/2,1/2)$.

Using the above calculation, we identify
\be
\biggl\langle \left(2\psi_1^{\left[\alpha \dot{A}\right.}\psi_2^{\left.\beta \dot{B}\right]}\right)(z_0)\, \sigma^{W\ti{X}}_{12}(z_1,\zb_1)\, \sigma^{Y\ti{Z}}_{12}(z_2,\zb_2)\biggr\rangle=\left(1/2[\Gamma^{\alpha\dot{A}},\Gamma^{\beta \dot{B}}]C\right)^{W\ti{X}Y\ti{Z}}\frac{1}{z_{01} z_{02}  \zb_{12}},
\ee
which may be verified by performing other calculations similar to the above.  The calculation when the indices $\alpha \dot{A}$ is the conjugate of $\beta \dot{B}$ are slightly different, but give the same result that is stated here.

If we interchange the locations of the two twist fields and their indices, i.e. $W\leftrightarrow Y$, $\ti{X}\leftrightarrow \ti{Z}$ and $z_1\leftrightarrow z_2$, we expect the answer to be even under this interchange.  Because of the antisymmetry of the gamma matrix $[\Gamma^{\alpha\dot{A}},\Gamma^{\beta \dot{B}}]C$, this is indeed the case. It is critically important that the cocycles gave the various signs to account for this antisymmetry.

Let us emphasize the relevance of the above calculation.  We have performed a computation which lifted to a four point function in the cover.  This four point function would be na\"{\i}vely disallowed if we restricted to the integral lattice of the last subsection.  This is because there were single fermions participating in the correlator.  However, the additional structure from the base space, namely the inherited radial ordering of fields, forces one to treat the two fields $\psi(t_+)$ and $\psi(t_-)$ as one unit.  One may not simply move one of the fermion operators past a spin field: one must move both to be consistent with the radial ordering in the $z$ plane, given that $\psi_1$ and $\psi_2$ are at the same point in the base.

We generalize the above computation for a general pair of twist $n$ fields, along with an untwisted single fermion insertion, in appendix \ref{app_3pf}.

\subsection{Twisted sector three-point functions}
There is one more situation where we might run into single fermions in the covering space in three point functions: when there is a fermionic excitation of a twist operator $\sigma_n$ with $n$ odd.  We consider a simple case where the twist operators are $\sigma_{(12)}$, $\sigma_{(23)}$ and $\sigma_{(123)}$.  To excite the $n=3$ twist operator, we act with modes of fermion fields. Thus, we will first have to discuss this excitation.

First, the bare twist $\sigma_{(123)}$ lifts to the identity operator on the cover: all of the information is contained in the conformal anomaly.  For simplicity, we place the operator $\sigma_{(123)}$ at the origin of the base space, and the the ramified image in the cover also at the origin.



We might be tempted to try
\be
\oint_0 \frac{dz}{2\pi i} z^{-2/3}(\psi^{\alpha \dot{A}}_1(z)+\omega \psi^{\alpha \dot{A}}_2(z)+\omega^{2}\psi^{\alpha \dot{A}}_3(z))\sigma_{(123)},
\ee
where $\omega=e^{2\pi i/3}$, which is a mode $\psi^{\alpha \dot{A}}_{-1/6}$ acting on the twist field.\footnote{Note, that if one changes $\omega\rightarrow \omega^2$ and $z^{-2/3}\rightarrow z^{-1/3}$ one again gets a sensible excitation from the standpoint of having a well defined integral.  However, this is an excitation $\psi_{1/6}$ mode which annihilates the above twist field, which can be easily verified.  This matches up with statements that the bare twist field should be the lowest weight in the given twist sector \cite{Lunin:2001pw,Lunin:2000yv}.}  However, this excitation will result in a spin $1/6$ operator.  Such an operator would have fractional statistics, and we wish to avoid such complications here\footnote{See \cite{Castro:2014tta} for holographic considerations along these lines, albeit for a system with unequal central charges in left and right moving sectors.  See also \cite{Lerda:1992ra} and references therein.}.  One could overcome this by acting with a right moving fermion to cancel the spin.  However, we find it more interesting to achieve $1/2$ integer spin by multiple applications of modes like the above.  To be concrete, we introduce the notation
\be
\psi^{\alpha \dot{A}}_{\omega^2}(z)=\psi^{\alpha \dot{A}}_1(z)+\omega^{}\psi^{\alpha \dot{A}}_2(z)+\omega^{2}\psi^{\alpha \dot{A}}_3(z)
\ee
and consider the excited twist field
\bea
\sigma_{(123)}^{+\dot{1}}=\oint_0 \frac{dz_3}{2\pi i} z_3^{-2/3}\psi_{\omega^2}^{-\dot{1}}(z_3)\oint_0 \frac{dz_2}{2\pi i} z_2^{-2/3}\psi_{\omega^2}^{+\dot{2}}(z_2)\oint_0 \frac{dz_1}{2\pi i} z_1^{-2/3}\psi_{\omega^2}^{+\dot{1}}(z_1)\sigma_{(123)}.
\eea
Note that the charges of the left two fermions cancel, and further that these two fermions share an OPE.  For this reason, we will need to consider the map to next to leading order.

This object lifts to the cover via the map $z=at^3+bt^4+\cdots$.  The first two maps lead to simple insertions in the $t$ plane of the corresponding fermions, along with factors of $a^{-1/6}\sqrt{3}$ for each excitation.  However, owing to the singularity with the next fermion, there is a possibility for multiple terms, and so we take the expansion of $z(t)$ and $dz/dt$ to next to leading order.  We find
\bea
\sigma_{(123)}^{+\dot{1}}&&\rightarrow \oint \frac{dt}{2\pi i} \left(\frac{dz}{dt}\right)^{1/2} z^{-2/3}\psi^{-\dot{1}}(t) a^{-1/3}:\psi^{+\dot{2}}(0)\psi^{+\dot{1}}(0): \\
&& = \oint \frac{dt}{2\pi i} \left(3at^2+4bt^3+{\mathcal{O}}(t^4)\right)^{1/2} (at^3+bt^4+{\mathcal{O}}(t^5))^{-2/3} \nn \\
&&\kern +3em \times a^{-1/3}\left(:\psi^{-\dot{1}}(t) \psi^{+\dot{2}}(0)\psi^{+\dot{1}}(0):-\frac{1}{t}\psi^{+\dot{2}}(0)\right) \nn \\
&& =\oint \frac{dt}{2\pi i} 3\sqrt{3}a^{-1/2}t^{-1}\left(1+{\mathcal{O}}(t^2)\right)\left(:\psi^{-\dot{1}}(t) \psi^{+\dot{2}}(0)\psi^{+\dot{1}}(0):-\frac{1}{t}\psi^{+\dot{2}}(0)\right),  \nn
\eea
and so, owing to some cancelation, we only get the leading order.  This gives
\bea
\sigma_{(123)}^{+\dot{1}}&\rightarrow & 3\sqrt{3}a^{-1/2}:\psi^{-\dot{1}} \psi^{+\dot{2}}\psi^{+\dot{1}}:.
\eea

We consider the three point function
\be
\langle \sigma_{(23)}^{-\dot{+}}(z_1,\zb_1) \sigma_{(12)}^{\dot{2}\dot{-}}(z_2,\zb_2)\sigma^{-\dot{2}}_{(123)}(z_3,\zb_3)\rangle.
\ee
The map from the base to the cover may be written
\be
\frac{z-z_1}{z-z_2}=\frac{(t-t_1)^2(2(t-t_2)t_{13}+(t-t_3)t_{12})}{(t-t_2)^2 (2(t-t_1)t_{23}-(t-t_3)t_{12})} \frac{t_{23}^2}{t_{13}^2}\frac{z_{13}}{z_{23}}.
\ee
Solving for $z$, and taking series around $t_1$ and $t_2$, and $t_3$, we find
\bea
&& (z-z_1)=-3\frac{t_{23}^2}{t_{13}^2t_{12}^2}\frac{z_{13}z_{12}}{z_{23}}(t-t_1)^2 +{\mathcal O}((t-t_1)^3),\\
&& (z-z_2)=3\frac{t_{13}^2}{t_{23}^2t_{12}^2}\frac{z_{23}z_{12}}{z_{13}}(t-t_2)^2 +{\mathcal O}((t-t_2)^3),\nn\\
&& (z-z_3)=-\frac{1}{2} \frac{t_{12}^3}{t_{23}^3t_{13}^3}\frac{z_{23}z_{13}}{z_{12}}(t-t_3)^3 + {\mathcal O}((t-t_3)^4).\nn
\eea
We also determine the location of the non-ramified images of $z_1$ and $z_2$.  These are given by
\be
\hat{t}_1=\frac{2t_2t_{13}+t_3t_{12}}{2t_{12}+t_{13}}, \qquad \hat{t}_2=\frac{2t_1t_{23}-t_3t_{12}}{-2t_{12}+t_{23}}.
\ee
We expand $z(t)$ around these points and find
\bea
(z-z_1)=-\frac{9}{8}\frac{(2t_{13}+t_{12})^2}{t_{12}t_{13}t_{23}} \frac{z_{12}z_{13}}{z_{23}}(t-\hat{t}_1)+{\mathcal O}((t-\hat{t}_1)^2),\\
(z-z_2)=-\frac{9}{8}\frac{(2t_{23}-t_{12})^2}{t_{12}t_{13}t_{23}} \frac{z_{12}z_{23}}{z_{13}}(t-\hat{t}_1)+{\mathcal O}((t-\hat{t}_2)^2).\nn
\eea

Using this, we lift the three point function to the covering surface
\bea
&&\langle \sigma_{23}^{-\dot{+}}(z_1,\zb_1) \sigma_{12}^{\dot{2}\dot{-}}(z_2,\zb_2)\sigma^{+\dot{1}}_{(123)}(z_3,\zb_3)\rangle \nn \\
&&\rightarrow \langle |b_1|^{-1/4} S^{-\dot{+}}_1 |b_2|^{-1/4}S^{\dot{2}\dot{-}}_2 3\sqrt{3}a^{-1/2}(\psi^{-\dot{1}} \psi^{+\dot{2}}\psi^{+\dot{1}})_3\rangle \nn \\
&&=9|t_{12}| \frac{1}{|z_{12}|^{1/2}} \sqrt{-2}\frac{t_{23}^{3/2}t_{13}^{3/2}}{t_{12}^{3/2}} \frac{z_{12}^{1/2}}{z_{23}^{1/2}z_{13}^{1/2}} \frac{([\Gamma^{+\dot{2}},\Gamma^{-\dot{1}}] \Gamma^{+\dot{1}}C)^{-\dot{+},\dot{2}\dot{-}}} {t_{13}^{1/2}t_{23}^{1/2}\tb_{12}^{1/2}}\frac{1}{2}\frac{-t_{12}}{t_{13}t_{23}} \nn \\
&&= -9\frac{\sqrt{-2}}{2|z_{12}|^{1/2}}\frac{z_{12}^{1/2}}{z_{23}^{1/2}z_{13}^{1/2}}([\Gamma^{+\dot{2}},\Gamma^{-\dot{1}}] \Gamma^{+\dot{1}}C)^{-\dot{+},\dot{2}\dot{-}}.\label{lifted3ptcorrelator}
\eea
Ignoring the overall proportionality, we multiply this by the functional dependence of the conformal anomaly, which in this case is $\frac{1}{|z_{12}|^{2/12}|z_{23}|^{4/3}|z_{13}|^{4/3}}$.  This gives the three point function
\be
\langle \sigma_{(23)}^{-\dot{+}}(z_1,\zb_1) \sigma_{(12)}^{\dot{2}\dot{-}}(z_2,\zb_2)\sigma^{+\dot{1}}_{(123)}(z_3,\zb_3)\rangle
\propto \frac{z_{12}^{1/6}}{z_{23}^{7/6}z_{13}^{7/6}} \frac{1}{\zb_{12}^{1/3}\zb_{23}^{2/3}\zb_{13}^{2/3}}([\Gamma^{+\dot{2}},\Gamma^{-\dot{1}}] \Gamma^{+\dot{1}}C)^{-\dot{+},\dot{2}\dot{-}}.
\ee
This agrees with the three point function for weights $(1/2,1/2)$, $(1/2,1/2)$, $(7/6,4/6)$, however, there are still branch cuts in the correlator.  We address this ambiguity now.

First, note that the above calculation is not the only calculation that one would have to do.  In the full calculation, one would have to perform an inner product over a sum of terms of the form
\be
\langle (\sigma_{(12)}+{\rm perm})(\sigma_{(12)}+{\rm perm})(\sigma_{(123)}+{\rm perm})\rangle,
\ee
where `${\rm perm}$' means to conjugate the indices with every group element of $S_{N}$ (here we suppress the charge indices).  Consider the cases where $\sigma_{(123)}$ or $\sigma_{(321)}$ is the operator chosen out of the twist three operator.  These will couple to any of the twist 2 operators that have indices in the set $1,2,3$. For example, both of the combinations of the form $\langle\sigma_{(23)}\sigma_{(12)}\sigma_{(123)}\rangle$ and $\langle\sigma_{(23)}\sigma_{(12)}\sigma_{(321)}\rangle$ appear.
However, these two can be related: if one takes the operator $\sigma_{(123)}$ and transports it around any of the twist two operators, it becomes $\sigma_{(321)}$.  Thus, we may relate these two contributions by taking $z_{23}\rightarrow e^{2\pi i}z_{23}$.  This results in a relative minus between the two terms, since there is a square root branch cut for $z_{12}$; this is most easily seen in equation (\ref{lifted3ptcorrelator}) because the conformal anomaly gives a real contribution, free of branch cuts. Hence, this pair of terms cancel. The rest of the terms can be organized in similar pairs, setting the full correlator to zero.

It appears that this is a general feature of correlators in the orbifold theory.  When computing individual correlators, corresponding to $S_{N}$ non-invariant pieces, one may find correlators that are ambiguous owing to branch cuts.  In these cases the full correlator has contributions from every branch, and so one sums over the roots of unity, giving a zero contribution.

The only cases left would be addressing correlators involving spins that were not integer or half integer values.  Usually one excludes such operators in a CFT because they are not local with respect to other operators.  We do not address such operators here.

Thus, restricting to integer or half integer spin, the rule seems to be that any correlator is acceptable.  If ambiguities owing to branch cuts are encountered, one sums over the branches, resulting in a zero correlator.  This may completely alleviate the need to place restrictions on the operators to arrive at a local operator algebra.  While the calculations of this section are not a proof, we feel that this is good evidence to suggest this claim.
\subsection{Deformation operator}

We now consider the deformation operator that deforms the theory away from the orbifold point.  This is given by a pair of supersymmetry generators acting on the twist two chiral primary, which we write
\be
O_{{A}{B}}=\frac{1}{4}\epsilon_{\alpha \beta}\epsilon_{\dot{\alpha}\dot{\beta}}\left[\oint_{z_0} \frac{dz}{2\pi i}G^{\alpha}_{\dot{A}}(z)\right]\left[\oint_{\zb_0} \frac{d\zb}{-2\pi i}\ti{G}^{\dot{\alpha}}_{{B}}(\zb)\right]\sigma_{2}^{\beta \dot{\beta}},
\ee
where $G^{\alpha}_{{A}}$ and $G^{\dot{\alpha}}_{{B}}$ are the supercurrents
\be
G^{\alpha}_{{A}}=\sum_i\psi_i^{\alpha \dot{A}}\pa X_{i,\dot{A}A},\qquad \ti{G}^{\dot{\alpha}}_{{B}}=\sum_i \ti{\psi}_i^{\dot{\alpha} \dot{A}} \pab X_{i,\dot{A} B},
\ee
and the twist field $\sigma_2^{\alpha\dot{\beta}}$ is the twist two operator $\sigma_{(12)}^{\alpha\dot{\beta}}$ plus all the images under $S_N$.  In \cite{Avery:2010er} it was shown that all four terms in the sums over $\alpha$ and $\dot{\beta}$ were proportional to the same operator/state by exploiting properties of the superconformal algebra and spectral flow.  Thus, the single term
\be
O_{{A}{B}}(z_0,\zb_0)=\left[\oint_{z_0} \frac{dz}{2\pi i}G^{-}_{{A}}(z)\right]\left[\oint_{\zb_0} \frac{d\zb}{-2\pi i}\ti{G}^{\dot{-}}_{{B}}(\zb)\right]\sigma_{2}^{+ \dot{+}}(z_0,\zb_0)
\ee
is a left and right $R$ symmetry singlet, and defines a family of four deformation operators.  These four operators can be grouped into a singlet and triplet of the $SU(2)_1$ symmetry.

For concreteness, we consider just one term when the copy indices are 1 and 2,
\be
\kern -2em O_{AB,(12)}(0,0)=\left[\oint \frac{dz}{2\pi i}\sum_{i=1}^2\psi_i^{- \dot{A}}\pa X_{i,\dot{A}A}(z)\right]\left[\oint \frac{d\zb}{-2\pi i}\sum_{i=1}^2 \ti{\psi}_i^{\dot{-} \dot{B}}\pab X_{i,\dot{B} B}(\zb)\right]\sigma_{(12)}^{+ \dot{+}}(0).
\ee
This lifts to the cover via the map $z=bt^2+\cdots$, giving
\bea
&& O_{AB,(12)}(0,0)\rightarrow  \\
&&\left[\oint \frac{dt}{2\pi i} \left(\frac{dz}{dt}\right)^{-1/2}\psi^{- \dot{A}}\pa X_{\dot{A}A}(z)\right]\left[\oint \frac{d\tb}{-2\pi i}\left(\frac{d\zb}{d\tb}\right)^{-1/2} \ti{\psi}^{\dot{-} \dot{B}}\pab X_{\dot{B} B}(\zb)\right]|b|^{-1/4}S^{+\dot{+}}.\nn
\eea
In the cover, only $\psi$ has a singular OPE with $S$.  The singularity is a $t^{-1/2}$, which combining with the $(dz/dt)^{-1/2}$ gives a simple pole, and so higher terms in the map are not needed.  The result is
\be
O_{AB,(12)}(0,0)\rightarrow |b|^{-5/4}\pa X_{\dot{A}A}\pab X_{\dot{B} B}\left(-\Gamma^{-\dot{A}}\ti{\Gamma}^{\dot{-}\dot{B}}\right)^{+\dot{+}}\,_{Y\ti{Z}}S^{Y\ti{Z}}.
\ee
One may check that all of the nonzero terms in the above gamma matrix structure are identical.  However, at this point we find it convenient to use a particular solution to the modulo arithmetic constraints.  We choose
\be
A_{56}=-1, A_{5\ti{5}}=-1, A_{5\ti{6}}=1, A_{6\ti{5}}=1, A_{6\ti{6}}=1, A_{\ti{5}\ti{6}}= 1.
\ee
With this,
\bea
O_{AB,(12)}(0,0)\rightarrow \hat{O}_{AB}=-i|b|^{-5/4}\bigg(&&\!\!\!\!\!\!\!\!\!\pa X_{\dot{1}A}\pab X_{\dot{1} B}S^{\dot{1}\dot{1}}+\pa X_{\dot{1}A}\pab X_{\dot{2} B}S^{\dot{1}\dot{2}} \nn \\
&&\!\!\!\!\!\!\!\!\!\!\!\!\!\!+\,\pa X_{\dot{2}A}\pab X_{\dot{1} B}S^{\dot{2}\dot{1}}+\pa X_{\dot{2}A}\pab X_{\dot{2} B}S^{\dot{2}\dot{2}} \bigg),
\eea
or
\bea
O_{AB,(12)}(0,0)\rightarrow \hat{O}_{AB}=-i|b|^{-5/4}\pa X_{\dot{A}A}\pab X_{\dot{B} B}S^{\dot{A}\dot{B}},
\eea
which makes it clear that the above operator is an $SU(2)_L\times SU(2)_R$ R-symmetry singlet, as well as being an $SU(2)_2$ singlet.
One may project this onto the singlet or the triplet of $SU(2)_1$ as one wishes.

The bosonization, along with the inclusion of the cocycles, makes certain features of the deformations evident.  For example, if we consider the singlet state
\be
O_{s,(12)}\equiv\epsilon^{AB}O_{AB,(12)}=O_{12,(12)}-O_{21,(12)}\rightarrow \epsilon^{AB}\hat{O}_{AB}\equiv \hat{O}_s,
\ee
one can show that it is self conjugate on the cover (as one might expect of a chargeless operator).  Further, the cocycles guarantee that the answer for the inner product of the singlet operator with itself is nonzero.  If one bosonizes, one would arrive at an answer of $\langle \hat{O}_{s}(t,\tb) \hat{O}_{s}(0,0)\rangle=0$; see \cite{Burrington:2012yq} where signs were put in by hand.

\section{Conclusions}

We have successfully written down a set of cocycles for the spin fields on the covering surface that are consistent with various $SU(2)$ symmetries in the theory, to help facilitate computations in the D1-D5 CFT near the orbifold point using the covering space techniques of \cite{Burrington:2012yq,Burrington:2012yn,Lunin:2001pw,Lunin:2000yv}.  We have checked that these cocycles give proper behavior for two point, three point, and certain four point functions in the cover. Further, we have considered some of the implications for three point functions in the base space, and shown how some problematic three point functions are avoided by inheriting radial ordering structure from the base space.

There are other ways of dealing with R sector operators; for example, in \cite{Burrington:2014yia,Carson:2014ena}, spectral flow is used to map the problem to one involving only NS sector operators (see \cite{Giusto:2004id,Lunin:2004uu} for applications in the dual gravity).  It would be interesting to see how the cocycles above reproduce results using spectral flow.  For example, one could consider the four point function of four spin fields, restricting the charges to be $+/-$ and $\dot{+}/\dot{-}$ so that spectral flow can change the periodicity around these points.  One would use spectral flow to change the periodicity of fermions at a pair of points where two spin fields are, changing them into NS sector operators, and then using a second spectral flow changing the periodicity of the fermions at the remaining two points where there are spin fields.  This would map the problem to an expectation value involving only NS sector operators.  Phases would be introduced during the two spectral flows because the other spin fields carry $SU(2)_R$ and $SU(2)_L$ R-charges. We believe these phases should reproduce the gamma matrix structure found for the four point function (\ref{4pointansatz}), although it would be interesting to check.

However, we should not lose track of the main goal of our current work, which is to facilitate the computation of four point functions using spin fields in the cover.  This is done so that we may track operator mixing in the presence of a deformation operator.  In our earlier work, \cite{Burrington:2012yq}, we showed that there is a simple state (which we call here) ${\mathcal O}_S$ with weight $h=1,\ti{h}=1$ in the untwisted sector that will mix with an operator of the same conformal weight (which we call here) ${\mathcal O}_{S'}$.  The new operator ${\mathcal O}_{S'}$ must be in the twist two sector.  This operator must be found, and shown to have the correct three point function with the deformation operator and the original operator ${\mathcal O}_S$.   After this, we must check to see if there are any fields of the same conformal dimension that ${\mathcal O}_{S'}$ mixes with, and so we must perform a four point function calculation with two deformation operators, and two ${\mathcal O}_{S'}$ operators.  Thus, the computation involving four twist two fields will lift to a covering space computation of four spin fields, which we have now gained control over computationally (\ref{4pointansatz}).

The mixing between two operators of the same conformal dimension introduces shifts in the conformal dimensions (see \cite{Dijkgraaf:1987jt} for a general discussion, and for similar computations in the D1-D5 CFT see \cite{Gaberdiel:2014cha}).  The entire block of fields with the same conformal dimension must be treated together to get the shift in the conformal dimensions.  Finding these shifts is an important step for interpolating between the `weak coupling' orbifold point and the `strong coupling' gravity description, and understanding which degrees of freedom are important for describing the Bekenstien-Hawking entropy for black holes away from extremality.  Exploring these shifts in conformal dimensions for states like ${\mathcal O}_S$ as well as twisted sector states is the problem to which we turn our attention in future work.

\section*{Acknowledgements}
We thank Samir Mathur for helpful discussions and Steve Avery for useful correspondence.
BAB is supported in part by a research grant from Hofstra University.
IGZ is supported in part by the U.S. National Science Foundation under CAREER grant no.~PHY10-53842, and by the U.S. Department of Energy under grant no.~DE-SC0009987.
AWP is supported in part by the Canadian Natural Sciences and Engineering Research Council under grant no.~SAPIN-2015-00034.

\appendix
\section{Bosonized symmetry currents and spin fields}
\subsection{\texorpdfstring{$SU(2)$}{} symmetry generators}\label{app_Js}

In this appendix we list the full bosonized form of the various $SU(2)$ symmetry generators of the theory dressed with appropriate cocycles. We find for the holomorphic R-symmetry
\bea
&& J^+=e^{i\pi M_{56}}e^{i\pi\left(M_{5i}-M_{6i}\right)\alpha_0^i}e^{i(\phi^5-\phi^6)},\\
&& J^-= e^{i\pi\left(-M_{55}-M_{66}+M_{65}\right)} e^{i\pi\left(-M_{5i}+M_{6i}\right)\alpha_0^i}e^{i(-\phi^5+\phi^6)},\\
&& J^3= \frac{i}{2}(\pa \phi^5-\pa \phi^6),
\eea
and for the antiholomorphic R-symmetry
\bea
&& \ti{J}^{\dot{+}}=e^{i\pi M_{\ti{5}\ti{6}}}e^{i\pi\left(M_{\ti{5}i}-M_{\ti{6}i}\right)\alpha_0^i}e^{i(\tp^{\ti{5}}-\tp^{\ti{6}})},\\
&& \ti{J}^{\dot{-}}= e^{i\pi\left(-M_{\ti{5}\ti{5}}-M_{\ti{6}\ti{6}}+M_{\ti{6}\ti{5}}\right)} e^{i\pi\left(-M_{\ti{5}i}+M_{\ti{6}i}\right)\alpha_0^i}e^{i(-\tp^{\ti{5}}+\tp^{\ti{6}})},\\
&& \ti{J}^{\dot{3}}= \frac{i}{2}(\pab \ti{\phi}^{\ti{5}}-\pab \ti{\phi}^{\ti{6}}).
\eea
For the holomorphic part of $SU(2)_2$ we find
\bea
&& {\mathcal{J}}^{\uparrow}=e^{i\pi\left(-M_{55}-M_{56}\right)}e^{i\pi\left(-M_{5i}-M_{6i}\right)\alpha_0^i}e^{i(-\phi^5-\phi^6)},\\
&& {\mathcal{J}}^{\downarrow}=e^{i\pi\left(-M_{65}-M_{66}\right)}e^{i\pi\left(M_{5i}+M_{6i}\right)\alpha_0^i} e^{i(\phi^5+\phi^6)},\\
&&{\mathcal J}^{0} = -\frac{i}{2}(\pa \phi^5+\pa \phi^6),
\eea
and for the antiholomorphic part of $SU(2)_2$ we find
\bea
&& \ti{\mathcal{J}}^{\uparrow}=e^{i\pi\left(-M_{\ti{5}\ti{5}}-M_{\ti{5}\ti{6}}\right)}e^{i\pi\left(-M_{\ti{5}i}-M_{\ti{6}i}\right)\alpha_0^i}e^{i(-\ti{\phi}^{\ti{5}}-\ti{\phi}^{\ti{6}})},\\
&& \ti{\mathcal{J}}^{\downarrow}=e^{i\pi\left(-M_{\ti{6}\ti{5}}-M_{\ti{6}\ti{6}}\right)}e^{i\pi\left(M_{\ti{5}i}+M_{\ti{6}i}\right)\alpha_0^i} e^{i(\ti{\phi}^{\ti{5}}+\ti{\phi}^{\ti{6}})},\\
&&\ti{\mathcal J}^{0} = -\frac{i}{2}(\pab \ti{\phi}^{\ti{5}}+\pab \ti{\phi}^{\ti{6}}).
\eea
As discussed below equation (\ref{Jcal0}), $SU(2)_2$ is really generated by the sum:
\be
{\mathbb J}^i= {\mathcal J}^i + \ti{\mathcal J}^i.
\ee
\subsection{Bosonized spin fields}\label{app_spinfields}
In this subsection we list the bosonized form of the spin fields which, according to the $R$-symmetry and internal $SU(2)$ indices they carry, are categorized into four families where each family is dressed with a phase factor of the form $e^{i\theta}$. We determine these phases explicitly in section \ref{thetaphases}. The analysis for the spin field $S^{+\dot+}$ and its SU(2) descendants has been described in detail in section \ref{spinfields}, and we summarize the result here:
\bea
S^{+\dot{+}}&=&e^{i\theta_{pp}}e^{\frac{i\pi}{2}\left(M_{5i}-M_{6i}+M_{\ti{5}i}-M_{\ti{6}i}\right)\alpha_0^i}
    e^{\frac{i}{2}(\phi^5-\phi^6+\tp^{\ti{5}}-\tp^{\ti{6}})},\\
S^{-\dot{+}}&=&e^{i\theta_{pp}}e^{\frac{i\pi}{2}\Big(
    \begin{smallmatrix}
    -M_{55}& +M_{65}& +M_{\ti{5}5}& - M_{\ti{6}5} \\
    -M_{56} & -M_{66} & -M_{\ti{5}6} & +M_{\ti{6}6}\end{smallmatrix}\Big)} e^{\frac{i\pi}{2}\left(-M_{5i}+M_{6i}+M_{\ti{5}i}-M_{\ti{6}i}\right)\alpha_0^i}
    e^{\frac{i}{2}\left(-\phi^5+\phi^6+\tp^{\ti{5}}-\tp^{\ti{6}}\right)},\\
S^{+\dot{-}}&=&e^{i\theta_{pp}}e^{\frac{i\pi}{2}\Big(
    \begin{smallmatrix}
    M_{5\ti{5}}& -M_{6\ti{5}}& -M_{\ti{5}\ti{5}}& + M_{\ti{6}\ti{5}} \\ -M_{5\ti{6}} & +M_{6\ti{6}} & -M_{\ti{5}\ti{6}} & -M_{\ti{6}\ti{6}} \end{smallmatrix}\Big)} e^{\frac{i\pi}{2}\left(M_{5i}-M_{6i}-M_{\ti{5}i}+M_{\ti{6}i}\right)\alpha_0^i}
    e^{\frac{i}{2}\left(\phi^5-\phi^6-\tp^{\ti{5}}+\tp^{\ti{6}}\right)},\\
S^{-\dot{-}}&=&e^{i\theta_{pp}}e^{\frac{i\pi}{2}\left(\begin{smallmatrix}
    -M_{55} & + M_{65} & + M_{\ti{5} 5} & - M_{\ti{6} 5} \\
    -M_{56} & - M_{66} & - M_{\ti{5} 6} & + M_{\ti{6} 6} \\
    -M_{5\ti{5}} & + M_{6\ti{5}} & - M_{\ti{5} \ti{5}} & + M_{\ti{6} \ti{5}} \\
    +M_{5\ti{6}} & - M_{6\ti{6}} & - M_{\ti{5} \ti{6}} & - M_{\ti{6} \ti{6}} \\
    \end{smallmatrix}
    \right)}e^{\frac{i\pi}{2}(-M_{5i}+M_{6i}-M_{\ti{5}i}+M_{\ti{6}i})\alpha_0^i} e^{\frac{i}{2}(-\phi^5+\phi^6-\tp^{\ti{5}}+\tp^{\ti{6}})}.
\eea
As explained below equation (\ref{Smp_detail}), for spin fields which carry an overall phase in terms of a sum over $M_{ab}$, we use a notation of writing this sum in two or four lines. This helps only with organizing various term in terms of their holomorphic and anti-holomorphic indices and does \emph{not} represent a matrix. The other three sectors are similarly obtained:  applying the modes of $J^{-}$ and $\ti{\mathcal J}^{\downarrow}$ to $S^{+\dot{1}}$ we find the set
\bea
S^{+\dot{1}}&=&e^{i\theta_{p1}}e^{\frac{i\pi}{2} \left(M_{5i}-M_{6i}-M_{\ti{5}i}-M_{\ti{6}i}\right)\alpha_0^i} e^{\frac{i}{2}\left(\phi^5-\phi^6-\tp^{\ti{5}}-\tp^{\ti{6}}\right)},\\
S^{- \dot{1}}&=&e^{i\theta_{p1}}e^{\frac{i\pi}{2}\left(\begin{smallmatrix}
-M_{55} & +M_{65} & -M_{\ti{5}5}& -M_{\ti{6}5} \\
-M_{56}& -M_{66} & +M_{\ti{5}6} & +M_{\ti{6}6} \end{smallmatrix}\right)}
e^{\frac{i\pi}{2}\left(-M_{5i}+M_{6i}-M_{\ti{5}i}-M_{\ti{6}i}\right)\alpha_0^i}
e^{\frac{i}{2}\left(-\phi^5+\phi^6-\tp^{\ti{5}}-\tp^{\ti{6}}\right)}, \\
S^{+\dot{2}}&=&e^{i\theta_{p1}}e^{\frac{i\pi}{2}\left(\begin{smallmatrix}
-M_{5\ti{5}} & +M_{6\ti{5}} & +M_{\ti{5}\ti{5}}& -M_{\ti{6}\ti{5}} \\
-M_{5\ti{6}}& +M_{6\ti{6}} & +M_{\ti{5}\ti{6}} & -M_{\ti{6}\ti{6}} \end{smallmatrix}\right)}
e^{\frac{i\pi}{2}\left(M_{5i}-M_{6i}+M_{\ti{5}i}+M_{\ti{6}i}\right)\alpha_0^i}
e^{\frac{i}{2}\left(\phi^5-\phi^6+\tp^{\ti{5}}+\tp^{\ti{6}}\right)}, \\
S^{-\dot{2}}&=&e^{i\theta_{p1}}e^{\frac{i\pi}{2}\left(\begin{smallmatrix}
-M_{55} & +M_{65} & -M_{\ti{5}5} & -M_{\ti{6}5} \\
-M_{56} & - M_{66} & + M_{\ti{5}6} & +M_{\ti{6}6} \\
+M_{5\ti{5}} & -M_{6\ti{5}} & +M_{\ti{5}\ti{5}}& -M_{\ti{6}\ti{5}} \\
+M_{5\ti{6}}& -M_{6\ti{6}} & +M_{\ti{5}\ti{6}} & -M_{\ti{6}\ti{6}} \end{smallmatrix}\right)}
e^{\frac{i\pi}{2}\left(-M_{5i}+M_{6i}+M_{\ti{5}i}+M_{\ti{6}i}\right)\alpha_0^i}
e^{\frac{i}{2}\left(-\phi^5+\phi^6+\tp^{\ti{5}}+\tp^{\ti{6}}\right)}.
\eea
The descendants of $S^{\dot{1}\dot{+}}$ are
\bea
S^{\dot{1}\dot{+}}&=&e^{i\theta_{1p}}e^{\frac{i\pi}{2} \left(-M_{5i}-M_{6i}+M_{\ti{5}i}-M_{\ti{6}i}\right)\alpha_0^i} e^{\frac{i}{2}\left(-\phi^5-\phi^6+\tp^{\ti{5}}-\tp^{\ti{6}}\right)},\\
S^{\dot{2}\dot{+}}&=&e^{i\theta_{1p}}e^{\frac{i\pi}{2}\left(\begin{smallmatrix}
M_{55} & -M_{65} & -M_{\ti{5}5}& +M_{\ti{6}5} \\
+M_{56}& -M_{66} & -M_{\ti{5}6} & +M_{\ti{6}6} \end{smallmatrix}\right)}
e^{\frac{i\pi}{2}\left(M_{5i}+M_{6i}+M_{\ti{5}i}-M_{\ti{6}i}\right)\alpha_0^i}
e^{\frac{i}{2}\left(\phi^5+\phi^6+\tp^{\ti{5}}-\tp^{\ti{6}}\right)}, \\
S^{\dot{1}\dot{-}}&=&e^{i\theta_{1p}}e^{\frac{i\pi}{2}\left(\begin{smallmatrix}
-M_{5\ti{5}} & -M_{6\ti{5}} & -M_{\ti{5}\ti{5}}& +M_{\ti{6}\ti{5}} \\
+M_{5\ti{6}}& +M_{6\ti{6}} & -M_{\ti{5}\ti{6}} & -M_{\ti{6}\ti{6}} \end{smallmatrix}\right)}
e^{\frac{i\pi}{2}\left(-M_{5i}-M_{6i}-M_{\ti{5}i}+M_{\ti{6}i}\right)\alpha_0^i}
e^{\frac{i}{2}\left(-\phi^5-\phi^6-\tp^{\ti{5}}+\tp^{\ti{6}}\right)}, \\
S^{\dot{2}\dot{-}}&=&e^{i\theta_{1p}}e^{\frac{i\pi}{2}\left(\begin{smallmatrix}
M_{55} & -M_{65} & +M_{\ti{5}5} & -M_{\ti{6}5} \\
+M_{56} & - M_{66} & + M_{\ti{5}6} & -M_{\ti{6}6} \\
-M_{5\ti{5}} & -M_{6\ti{5}} & -M_{\ti{5}\ti{5}}& +M_{\ti{6}\ti{5}} \\
+M_{5\ti{6}}& +M_{6\ti{6}} & -M_{\ti{5}\ti{6}} & -M_{\ti{6}\ti{6}} \end{smallmatrix}\right)}
e^{\frac{i\pi}{2}\left(M_{5i}+M_{6i}-M_{\ti{5}i}+M_{\ti{6}i}\right)\alpha_0^i}
e^{\frac{i}{2}\left(\phi^5+\phi^6-\tp^{\ti{5}}+\tp^{\ti{6}}\right)};
\eea
and the descendants of $S^{\dot{1}\dot{1}}$ are
\bea
S^{\dot{1}\dot{1}}&=&e^{i\theta_{11}}e^{\frac{i\pi}{2} \left(-M_{5i}-M_{6i}-M_{\ti{5}i}-M_{\ti{6}i}\right)\alpha_0^i} e^{\frac{i}{2}\left(-\phi^5-\phi^6-\tp^{\ti{5}}-\tp^{\ti{6}}\right)},\\
S^{\dot{2}\dot{1}}&=&e^{i\theta_{11}}e^{\frac{i\pi}{2}\left(\begin{smallmatrix}
M_{55} & -M_{65} & +M_{\ti{5}5}& +M_{\ti{6}5} \\
+M_{56}& -M_{66} & +M_{\ti{5}6} & +M_{\ti{6}6} \end{smallmatrix}\right)}
e^{\frac{i\pi}{2}\left(M_{5i}+M_{6i}-M_{\ti{5}i}-M_{\ti{6}i}\right)\alpha_0^i}
e^{\frac{i}{2}\left(\phi^5+\phi^6-\tp^{\ti{5}}-\tp^{\ti{6}}\right)}, \\
S^{\dot{1}\dot{2}}&=&e^{i\theta_{11}}e^{\frac{i\pi}{2}\left(\begin{smallmatrix}
M_{5\ti{5}} & +M_{6\ti{5}} & +M_{\ti{5}\ti{5}}& -M_{\ti{6}\ti{5}} \\
+M_{5\ti{6}}& +M_{6\ti{6}} & +M_{\ti{5}\ti{6}} & -M_{\ti{6}\ti{6}} \end{smallmatrix}\right)}
e^{\frac{i\pi}{2}\left(-M_{5i}-M_{6i}+M_{\ti{5}i}+M_{\ti{6}i}\right)\alpha_0^i}
e^{\frac{i}{2}\left(-\phi^5-\phi^6+\tp^{\ti{5}}+\tp^{\ti{6}}\right)}, \\
S^{\dot{2}\dot{2}}&=&e^{i\theta_{11}}e^{\frac{i\pi}{2}\left(\begin{smallmatrix}
M_{55} & -M_{65} & +M_{\ti{5}5} & +M_{\ti{6}5} \\
+M_{56} & - M_{66} & + M_{\ti{5}6} & +M_{\ti{6}6} \\
-M_{5\ti{5}} & -M_{6\ti{5}} & +M_{\ti{5}\ti{5}}& -M_{\ti{6}\ti{5}} \\
-M_{5\ti{6}}& -M_{6\ti{6}} & +M_{\ti{5}\ti{6}} & -M_{\ti{6}\ti{6}} \end{smallmatrix}\right)}
e^{\frac{i\pi}{2}\left(M_{5i}+M_{6i}+M_{\ti{5}i}+M_{\ti{6}i}\right)\alpha_0^i}
e^{\frac{i}{2}\left(\phi^5+\phi^6+\tp^{\ti{5}}+\tp^{\ti{6}}\right)}.
\eea
\section{Gamma matrices}

\label{gammaappendix}
Rather than writing these out as $16\times16$ matrices, with most of the terms being zero, we will simply state the non-zero terms.  Below, we write these before the choice of phases dressing the various spin fields (see appendix \ref{app_spinfields}).
\bea
&& \psi^{-\dot{1}}:\nn \\
&& (\Gamma^{-\dot{1}})^{+\dot{+}}\,_{\dot{1} \dot{+}}=e^{i(\theta_{pp}-\theta_{1p})} e^{\frac{i\pi}{2}(-M_{55}-M_{65}+M_{\ti{5}5}-M_{\ti{6}5})}, \nn \\
&& (\Gamma^{-\dot{1}})^{+\dot{-}}\,_{\dot{1} \dot{-}}=e^{i(\theta_{pp}-\theta_{1p})} e^{\frac{i\pi}{2}(-M_{55}-M_{65}+M_{\ti{5}5}-M_{\ti{6}5})}, \nn \\
&& (\Gamma^{-\dot{1}})^{+\dot{1}}\,_{\dot{1} \dot{1}}=e^{i(\theta_{p1}-\theta_{11})} e^{\frac{i\pi}{2}(-M_{55}-M_{65}-M_{\ti{5}5}-M_{\ti{6}5})}, \nn \\
&& (\Gamma^{-\dot{1}})^{+\dot{2}}\,_{\dot{1} \dot{2}}=e^{i(\theta_{p1}-\theta_{11})} e^{\frac{i\pi}{2}(-M_{55}-M_{65}-M_{\ti{5}5}-M_{\ti{6}5})}, \label{m1phases}\\
&& (\Gamma^{-\dot{1}})^{\dot{2}\dot{+}}
\,_{- \dot{+}}=e^{i(\theta_{1p}-\theta_{pp})} e^{\frac{i\pi}{2}(M_{55}-M_{65}-M_{\ti{5}5}+M_{\ti{6}5}+2M_{56})}, \nn \\
&& (\Gamma^{-\dot{1}})^{\dot{2}\dot{-}}
\,_{- \dot{-}}=e^{i(\theta_{1p}-\theta_{pp})} e^{\frac{i\pi}{2}(M_{55}-M_{65}-M_{\ti{5}5}+M_{\ti{6}5}+2M_{56})}, \nn \\
&& (\Gamma^{-\dot{1}})^{\dot{2}\dot{1}}
\,_{- \dot{1}}=e^{i(\theta_{11}-\theta_{p1})} e^{\frac{i\pi}{2}(M_{55}-M_{65}+M_{\ti{5}5}+M_{\ti{6}5}+2M_{56})},\nn \\
&& (\Gamma^{-\dot{1}})^{\dot{2}\dot{2}}
\,_{- \dot{2}}=e^{i(\theta_{11}-\theta_{p1})} e^{\frac{i\pi}{2}(M_{55}-M_{65}+M_{\ti{5}5}+M_{\ti{6}5}+2M_{56})},\nn
\eea

\bea
&& \psi^{+\dot{2}}:\nn \\
&& (\Gamma^{+\dot{2}})^{-\dot{+}}
\,_{\dot{2} \dot{+}}=e^{i(\theta_{pp}-\theta_{1p})} e^{\frac{i\pi}{2}(-M_{55}+M_{65}+M_{\ti{5}5}-M_{\ti{6}5}-2M_{56})}, \nn \\
&& (\Gamma^{+\dot{2}})^{-\dot{-}}
\,_{\dot{2} \dot{-}}=e^{i(\theta_{pp}-\theta_{1p})} e^{\frac{i\pi}{2}(-M_{55}+M_{65}+M_{\ti{5}5}-M_{\ti{6}5}-2M_{56})}, \nn \\
&& (\Gamma^{+\dot{2}})^{-\dot{1}}
\,_{\dot{2} \dot{1}}=e^{i(\theta_{p1}-\theta_{11})} e^{\frac{i\pi}{2}(-M_{55}+M_{65}-M_{\ti{5}5}-M_{\ti{6}5}-2M_{56})}, \nn \\
&& (\Gamma^{+\dot{2}})^{-\dot{2}}
\,_{\dot{2} \dot{2}}=e^{i(\theta_{p1}-\theta_{11})} e^{\frac{i\pi}{2}(-M_{55}+M_{65}-M_{\ti{5}5}-M_{\ti{6}5}-2M_{56})},  \label{p2phases}\\
&& (\Gamma^{+\dot{2}})^{\dot{1}\dot{+}}
\,_{+ \dot{+}}=e^{i(\theta_{1p}-\theta_{pp})} e^{\frac{i\pi}{2}(M_{55}+M_{65}-M_{\ti{5}5}+M_{\ti{6}5})}, \nn \\
&& (\Gamma^{+\dot{2}})^{\dot{1}\dot{-}}
\,_{+ \dot{-}}=e^{i(\theta_{1p}-\theta_{pp})} e^{\frac{i\pi}{2}(M_{55}+M_{65}-M_{\ti{5}5}+M_{\ti{6}5})}, \nn \\
&& (\Gamma^{+\dot{2}})^{\dot{1}\dot{1}}
\,_{+ \dot{1}}=e^{i(\theta_{11}-\theta_{p1})} e^{\frac{i\pi}{2}(M_{55}+M_{65}+M_{\ti{5}5}+M_{\ti{6}5})}, \nn \\
&& (\Gamma^{+\dot{2}})^{\dot{1}\dot{2}}
\,_{+ \dot{2}}=e^{i(\theta_{11}-\theta_{p1})} e^{\frac{i\pi}{2}(M_{55}+M_{65}+M_{\ti{5}5}+M_{\ti{6}5})}, \nn
\eea

\bea
&& \psi^{-\dot{2}}:\nn \\
&& (\Gamma^{-\dot{2}})^{+\dot{+}}
\,_{\dot{2} \dot{+}}=-e^{i(\theta_{pp}-\theta_{1p})} e^{\frac{i\pi}{2}(-M_{55}+M_{65}+M_{\ti{5}5}-M_{\ti{6}5}-2M_{56})}, \nn \\
&& (\Gamma^{-\dot{2}})^{+\dot{-}}
\,_{\dot{2} \dot{-}}=-e^{i(\theta_{pp}-\theta_{1p})} e^{\frac{i\pi}{2}(-M_{55}+M_{65}+M_{\ti{5}5}-M_{\ti{6}5}-2M_{56})}, \nn \\
&& (\Gamma^{-\dot{2}})^{+\dot{1}}
\,_{\dot{2} \dot{1}}=-e^{i(\theta_{p1}-\theta_{11})} e^{\frac{i\pi}{2}(-M_{55}+M_{65}-M_{\ti{5}5}-M_{\ti{6}5}-2M_{56})}, \nn \\
&& (\Gamma^{-\dot{2}})^{+\dot{2}}
\,_{\dot{2} \dot{2}}=-e^{i(\theta_{p1}-\theta_{11})} e^{\frac{i\pi}{2}(-M_{55}+M_{65}-M_{\ti{5}5}-M_{\ti{6}5}-2M_{56})}, \label{m2phases} \\
&& (\Gamma^{-\dot{2}})^{\dot{1}\dot{+}}
\,_{{-} \dot{+}}=-e^{i(\theta_{1p}-\theta_{pp})} e^{\frac{i\pi}{2}(M_{55}-M_{65}-M_{\ti{5}5}+M_{\ti{6}5}+2M_{56})}, \nn \\
&& (\Gamma^{-\dot{2}})^{\dot{1}\dot{-}}
\,_{{-} \dot{-}}=-e^{i(\theta_{1p}-\theta_{pp})} e^{\frac{i\pi}{2}(M_{55}-M_{65}-M_{\ti{5}5}+M_{\ti{6}5}+2M_{56})}, \nn \\
&& (\Gamma^{-\dot{2}})^{\dot{1}\dot{1}}
\,_{{-} \dot{1}}=-e^{i(\theta_{11}-\theta_{p1})} e^{\frac{i\pi}{2}(M_{55}-M_{65}+M_{\ti{5}5}+M_{\ti{6}5}+2M_{56})}, \nn \\
&& (\Gamma^{-\dot{2}})^{\dot{1}\dot{2}}
\,_{{-} \dot{2}}=-e^{i(\theta_{11}-\theta_{p1})} e^{\frac{i\pi}{2}(M_{55}-M_{65}+M_{\ti{5}5}+M_{\ti{6}5}+2M_{56})}, \nn
\eea

\bea
&& \psi^{+\dot{1}}:\nn \\
&& (\Gamma^{+\dot{1}})^{-\dot{+}}
\,_{\dot{1} \dot{+}}=e^{i(\theta_{pp}-\theta_{1p})} e^{\frac{i\pi}{2}(-M_{55}+M_{65}+M_{\ti{5}5}-M_{\ti{6}5}-2M_{56})}, \nn \\
&& (\Gamma^{+\dot{1}})^{-\dot{-}}
\,_{\dot{1} \dot{-}}=e^{i(\theta_{pp}-\theta_{1p})} e^{\frac{i\pi}{2}(-M_{55}+M_{65}+M_{\ti{5}5}-M_{\ti{6}5}-2M_{56})}, \nn \\
&& (\Gamma^{+\dot{1}})^{-\dot{1}}
\,_{\dot{1} \dot{1}}=e^{i(\theta_{p1}-\theta_{11})} e^{\frac{i\pi}{2}(-M_{55}+M_{65}-M_{\ti{5}5}-M_{\ti{6}5}-2M_{56})}, \nn \\
&& (\Gamma^{+\dot{1}})^{-\dot{2}}
\,_{\dot{1} \dot{2}}=e^{i(\theta_{p1}-\theta_{11})} e^{\frac{i\pi}{2}(-M_{55}+M_{65}-M_{\ti{5}5}-M_{\ti{6}5}-2M_{56})}, \label{p1phases} \\
&& (\Gamma^{+\dot{1}})^{\dot{2}\dot{+}}
\,_{+ \dot{+}}=e^{i(\theta_{1p}-\theta_{pp})} e^{\frac{i\pi}{2}(M_{55}-M_{65}-M_{\ti{5}5}+M_{\ti{6}5}+2M_{56})}, \nn \\
&& (\Gamma^{+\dot{1}})^{\dot{2}\dot{-}}
\,_{+ \dot{-}}=e^{i(\theta_{1p}-\theta_{pp})} e^{\frac{i\pi}{2}(M_{55}-M_{65}-M_{\ti{5}5}+M_{\ti{6}5}+2M_{56})}, \nn \\
&& (\Gamma^{+\dot{1}})^{\dot{2}\dot{1}}
\,_{+ \dot{1}}=e^{i(\theta_{11}-\theta_{p1})} e^{\frac{i\pi}{2}(M_{55}-M_{65}+M_{\ti{5}5}+M_{\ti{6}5}+2M_{56})}, \nn \\
&& (\Gamma^{+\dot{1}})^{\dot{2}\dot{2}}
\,_{+ \dot{2}}=e^{i(\theta_{11}-\theta_{p1})} e^{\frac{i\pi}{2}(M_{55}-M_{65}+M_{\ti{5}5}+M_{\ti{6}5}+2M_{56})}. \nn
\eea
For the antiholomorphic fermions we find
\bea
&& \ti{\psi}^{\dot{-}\dot{1}}:\nn \\
&& (\ti{\Gamma}^{\dot{-}\dot{1}})^{+\dot{+}}
\,_{+\dot{1}}=e^{i(\theta_{pp}-\theta_{p1})} e^{\frac{i\pi}{2}(M_{5\ti{5}}-M_{6\ti{5}}-M_{\ti{5}\ti{5}}-M_{\ti{6}\ti{5}})}, \nn \\
&& (\ti{\Gamma}^{\dot{-}\dot{1}})^{-\dot{+}}
\,_{-\dot{1}}=e^{i(\theta_{pp}-\theta_{p1})} e^{\frac{i\pi}{2}(M_{5\ti{5}}-M_{6\ti{5}}-M_{\ti{5}\ti{5}}-M_{\ti{6}\ti{5}})}, \nn \\
&& (\ti{\Gamma}^{\dot{-}\dot{1}})^{\dot{1}\dot{+}}
\,_{\dot{1}\dot{1}}=e^{i(\theta_{1p}-\theta_{11})} e^{\frac{i\pi}{2}(-M_{5\ti{5}}-M_{6\ti{5}}-M_{\ti{5}\ti{5}}-M_{\ti{6}\ti{5}})}, \nn \\
&& (\ti{\Gamma}^{\dot{-}\dot{1}})^{\dot{2}\dot{+}}
\,_{\dot{2}\dot{1}}=e^{i(\theta_{1p}-\theta_{11})} e^{\frac{i\pi}{2}(-M_{5\ti{5}}-M_{6\ti{5}}-M_{\ti{5}\ti{5}}-M_{\ti{6}\ti{5}})}, \\
&& (\ti{\Gamma}^{\dot{-}\dot{1}})^{+\dot{2}}
\,_{+\dot{-}}=e^{i(\theta_{p1}-\theta_{pp})} e^{\frac{i\pi}{2}(-M_{5\ti{5}}+M_{6\ti{5}}+M_{\ti{5}\ti{5}}-M_{\ti{6}\ti{5}}+2M_{\ti{5}\ti{6}})}, \nn \\
&& (\ti{\Gamma}^{\dot{-}\dot{1}})^{-\dot{2}}
\,_{-\dot{-}}=e^{i(\theta_{p1}-\theta_{pp})} e^{\frac{i\pi}{2}(-M_{5\ti{5}}+M_{6\ti{5}}+M_{\ti{5}\ti{5}}-M_{\ti{6}\ti{5}}+2M_{\ti{5}\ti{6}})}, \nn \\
&& (\ti{\Gamma}^{\dot{-}\dot{1}})^{\dot{1}\dot{2}}
\,_{\dot{1}\dot{-}}=e^{i(\theta_{11}-\theta_{1p})} e^{\frac{i\pi}{2}(M_{5\ti{5}}+M_{6\ti{5}}+M_{\ti{5}\ti{5}}-M_{\ti{6}\ti{5}}+2M_{\ti{5}\ti{6}})}, \nn \\
&& (\ti{\Gamma}^{\dot{-}\dot{1}})^{\dot{2}\dot{2}}
\,_{\dot{2}\dot{-}}=e^{i(\theta_{11}-\theta_{1p})} e^{\frac{i\pi}{2}(M_{5\ti{5}}+M_{6\ti{5}}+M_{\ti{5}\ti{5}}-M_{\ti{6}\ti{5}}+2M_{\ti{5}\ti{6}})}, \nn
\eea
\bea
&& \ti{\psi}^{\dot{+}\dot{2}}:\nn \\
&& (\ti{\Gamma}^{\dot{+}\dot{2}})^{+\dot{-}}
\,_{+\dot{2}}=e^{i(\theta_{pp}-\theta_{p1})} e^{\frac{i\pi}{2}(M_{5\ti{5}}-M_{6\ti{5}}-M_{\ti{5}\ti{5}}+M_{\ti{6}\ti{5}}-2M_{\ti{5}\ti{6}})}, \nn \\
&& (\ti{\Gamma}^{\dot{+}\dot{2}})^{-\dot{-}}
\,_{-\dot{2}}=e^{i(\theta_{pp}-\theta_{p1})} e^{\frac{i\pi}{2}(M_{5\ti{5}}-M_{6\ti{5}}-M_{\ti{5}\ti{5}}+M_{\ti{6}\ti{5}}-2M_{\ti{5}\ti{6}})}, \nn \\
&& (\ti{\Gamma}^{\dot{+}\dot{2}})^{\dot{1}\dot{-}}
\,_{\dot{1}\dot{2}}=e^{i(\theta_{1p}-\theta_{11})} e^{\frac{i\pi}{2}(-M_{5\ti{5}}-M_{6\ti{5}}-M_{\ti{5}\ti{5}}+M_{\ti{6}\ti{5}}-2M_{\ti{5}\ti{6}})}, \nn \\
&& (\ti{\Gamma}^{\dot{+}\dot{2}})^{\dot{2}\dot{-}}
\,_{\dot{2}\dot{2}}=e^{i(\theta_{1p}-\theta_{11})} e^{\frac{i\pi}{2}(-M_{5\ti{5}}-M_{6\ti{5}}-M_{\ti{5}\ti{5}}+M_{\ti{6}\ti{5}}-2M_{\ti{5}\ti{6}})}, \\
&& (\ti{\Gamma}^{\dot{+}\dot{2}})^{+\dot{1}}
\,_{+\dot{+}}=e^{i(\theta_{p1}-\theta_{pp})} e^{\frac{i\pi}{2}(-M_{5\ti{5}}+M_{6\ti{5}}+M_{\ti{5}\ti{5}}+M_{\ti{6}\ti{5}})}, \nn \\
&& (\ti{\Gamma}^{\dot{+}\dot{2}})^{-\dot{1}}
\,_{-\dot{+}}=e^{i(\theta_{p1}-\theta_{pp})} e^{\frac{i\pi}{2}(-M_{5\ti{5}}+M_{6\ti{5}}+M_{\ti{5}\ti{5}}+M_{\ti{6}\ti{5}})}, \nn \\
&& (\ti{\Gamma}^{\dot{+}\dot{2}})^{\dot{1}\dot{1}}
\,_{\dot{1}\dot{+}}=e^{i(\theta_{11}-\theta_{1p})} e^{\frac{i\pi}{2}(M_{5\ti{5}}+M_{6\ti{5}}+M_{\ti{5}\ti{5}}+M_{\ti{6}\ti{5}})}, \nn \\
&& (\ti{\Gamma}^{\dot{+}\dot{2}})^{\dot{2}\dot{1}}
\,_{\dot{2}\dot{+}}=e^{i(\theta_{11}-\theta_{1p})} e^{\frac{i\pi}{2}(M_{5\ti{5}}+M_{6\ti{5}}+M_{\ti{5}\ti{5}}+M_{\ti{6}\ti{5}})}, \nn
\eea

\bea
&& \ti{\psi}^{\dot{-}\dot{2}}:\nn \\
&& (\ti{\Gamma}^{\dot{-}\dot{2}})^{+\dot{+}}
\,_{+\dot{2}}=-e^{i(\theta_{pp}-\theta_{p1})} e^{\frac{i\pi}{2}(M_{5\ti{5}}-M_{6\ti{5}}-M_{\ti{5}\ti{5}}+M_{\ti{6}\ti{5}}-2M_{\ti{5}\ti{6}})}, \nn \\
&& (\ti{\Gamma}^{\dot{-}\dot{2}})^{-\dot{+}}
\,_{-\dot{2}}=-e^{i(\theta_{pp}-\theta_{p1})} e^{\frac{i\pi}{2}(M_{5\ti{5}}-M_{6\ti{5}}-M_{\ti{5}\ti{5}}+M_{\ti{6}\ti{5}}-2M_{\ti{5}\ti{6}})}, \nn \\
&& (\ti{\Gamma}^{\dot{-}\dot{2}})^{\dot{1}\dot{+}}
\,_{\dot{1}\dot{2}}=-e^{i(\theta_{1p}-\theta_{11})} e^{\frac{i\pi}{2}(-M_{5\ti{5}}-M_{6\ti{5}}-M_{\ti{5}\ti{5}}+M_{\ti{6}\ti{5}}-2M_{\ti{5}\ti{6}})}, \nn \\
&& (\ti{\Gamma}^{\dot{-}\dot{2}})^{\dot{2}\dot{+}}
\,_{\dot{2}\dot{2}}=-e^{i(\theta_{1p}-\theta_{11})} e^{\frac{i\pi}{2}(-M_{5\ti{5}}-M_{6\ti{5}}-M_{\ti{5}\ti{5}}+M_{\ti{6}\ti{5}}-2M_{\ti{5}\ti{6}})}, \\
&& (\ti{\Gamma}^{\dot{-}\dot{2}})^{+\dot{1}}
\,_{+\dot{-}}=-e^{i(\theta_{p1}-\theta_{pp})} e^{\frac{i\pi}{2}(-M_{5\ti{5}}+M_{6\ti{5}}+M_{\ti{5}\ti{5}}-M_{\ti{6}\ti{5}}+2M_{\ti{5}\ti{6}})}, \nn \\
&& (\ti{\Gamma}^{\dot{-}\dot{2}})^{-\dot{1}}
\,_{-\dot{-}}=-e^{i(\theta_{p1}-\theta_{pp})} e^{\frac{i\pi}{2}(-M_{5\ti{5}}+M_{6\ti{5}}+M_{\ti{5}\ti{5}}-M_{\ti{6}\ti{5}}+2M_{\ti{5}\ti{6}})}, \nn \\
&& (\ti{\Gamma}^{\dot{-}\dot{2}})^{\dot{1}\dot{1}}
\,_{\dot{1}\dot{-}}=-e^{i(\theta_{11}-\theta_{1p})} e^{\frac{i\pi}{2}(M_{5\ti{5}}+M_{6\ti{5}}+M_{\ti{5}\ti{5}}-M_{\ti{6}\ti{5}}+2M_{\ti{5}\ti{6}})}, \nn \\
&& (\ti{\Gamma}^{\dot{-}\dot{2}})^{\dot{2}\dot{1}}
\,_{\dot{2}\dot{-}}=-e^{i(\theta_{11}-\theta_{1p})} e^{\frac{i\pi}{2}(M_{5\ti{5}}+M_{6\ti{5}}+M_{\ti{5}\ti{5}}-M_{\ti{6}\ti{5}}+2M_{\ti{5}\ti{6}})}, \nn
\eea
\bea
&& \ti{\psi}^{\dot{+}\dot{1}}:\nn \\
&& (\ti{\Gamma}^{\dot{+}\dot{1}})^{+\dot{-}}
\,_{+\dot{1}}=e^{i(\theta_{pp}-\theta_{p1})} e^{\frac{i\pi}{2}(M_{5\ti{5}}-M_{6\ti{5}}-M_{\ti{5}\ti{5}}+M_{\ti{6}\ti{5}}-2M_{\ti{5}\ti{6}})}, \nn \\
&& (\ti{\Gamma}^{\dot{+}\dot{1}})^{-\dot{-}}
\,_{-\dot{1}}=e^{i(\theta_{pp}-\theta_{p1})} e^{\frac{i\pi}{2}(M_{5\ti{5}}-M_{6\ti{5}}-M_{\ti{5}\ti{5}}+M_{\ti{6}\ti{5}}-2M_{\ti{5}\ti{6}})}, \nn \\
&& (\ti{\Gamma}^{\dot{+}\dot{1}})^{\dot{1}\dot{-}}
\,_{\dot{1}\dot{1}}=e^{i(\theta_{1p}-\theta_{11})} e^{\frac{i\pi}{2}(-M_{5\ti{5}}-M_{6\ti{5}}-M_{\ti{5}\ti{5}}+M_{\ti{6}\ti{5}}-2M_{\ti{5}\ti{6}})}, \nn \\
&& (\ti{\Gamma}^{\dot{+}\dot{1}})^{\dot{2}\dot{-}}
\,_{\dot{2}\dot{1}}=e^{i(\theta_{1p}-\theta_{11})} e^{\frac{i\pi}{2}(-M_{5\ti{5}}-M_{6\ti{5}}-M_{\ti{5}\ti{5}}+M_{\ti{6}\ti{5}}-2M_{\ti{5}\ti{6}})}, \\
&& (\ti{\Gamma}^{\dot{+}\dot{1}})^{+\dot{2}}
\,_{+\dot{+}}=e^{i(\theta_{p1}-\theta_{pp})} e^{\frac{i\pi}{2}(-M_{5\ti{5}}+M_{6\ti{5}}+M_{\ti{5}\ti{5}}-M_{\ti{6}\ti{5}}+2M_{\ti{5}\ti{6}})}, \nn \\
&& (\ti{\Gamma}^{\dot{+}\dot{1}})^{-\dot{2}}
\,_{-\dot{+}}=e^{i(\theta_{p1}-\theta_{pp})} e^{\frac{i\pi}{2}(-M_{5\ti{5}}+M_{6\ti{5}}+M_{\ti{5}\ti{5}}-M_{\ti{6}\ti{5}}+2M_{\ti{5}\ti{6}})}, \nn \\
&& (\ti{\Gamma}^{\dot{+}\dot{1}})^{\dot{1}\dot{2}}
\,_{\dot{1}\dot{+}}=e^{i(\theta_{11}-\theta_{1p})} e^{\frac{i\pi}{2}(M_{5\ti{5}}+M_{6\ti{5}}+M_{\ti{5}\ti{5}}-M_{\ti{6}\ti{5}}+2M_{\ti{5}\ti{6}})}, \nn \\
&& (\ti{\Gamma}^{\dot{+}\dot{1}})^{\dot{2}\dot{2}}
\,_{\dot{2}\dot{+}}=e^{i(\theta_{11}-\theta_{1p})} e^{\frac{i\pi}{2}(M_{5\ti{5}}+M_{6\ti{5}}+M_{\ti{5}\ti{5}}-M_{\ti{6}\ti{5}}+2M_{\ti{5}\ti{6}})}. \nn
\eea
\section{Spin fields and the choice of the phases}\label{app_theta_spinfields}

In this appendix we list the complete bosonized form of the spin fields with the dressing phases derived in (\ref{thetapp})-(\ref{theta1p}):
\bea
\kern -2em S^{+\dot{+}}&=&i e^{\frac{i\pi}{8}\left(\begin{smallmatrix}
    +M_{55} & - M_{65} & - 3M_{\ti{5} 5} & + 3M_{\ti{6} 5} \\
    +3M_{56} & + M_{66} & + 3M_{\ti{5} 6} & - 3M_{\ti{6} 6} \\
    +M_{5\ti{5}} & - M_{6\ti{5}} & + M_{\ti{5} \ti{5}} & - M_{\ti{6} \ti{5}} \\
    -M_{5\ti{6}} & + M_{6\ti{6}} & + 3M_{\ti{5} \ti{6}} & + M_{\ti{6} \ti{6}} \\
    \end{smallmatrix}
    \right)}
    e^{\frac{i\pi}{2}\left(M_{5i}-M_{6i}+M_{\ti{5}i}-M_{\ti{6}i}\right)\alpha_0^i}
    e^{\frac{i}{2}(\phi^5-\phi^6+\tp^{\ti{5}}-\tp^{\ti{6}})},\\
\kern -2em S^{-\dot{+}}&=&i e^{\frac{i\pi}{8}\left(\begin{smallmatrix}
    -3M_{55} & + 3M_{65} & + M_{\ti{5} 5} & - M_{\ti{6} 5} \\
    -M_{56} & - 3M_{66} & - M_{\ti{5} 6} & + M_{\ti{6} 6} \\
    +M_{5\ti{5}} & - M_{6\ti{5}} & + M_{\ti{5} \ti{5}} & - M_{\ti{6} \ti{5}} \\
    -M_{5\ti{6}} & + M_{6\ti{6}} & + 3M_{\ti{5} \ti{6}} & + M_{\ti{6} \ti{6}} \\
    \end{smallmatrix}
    \right)}
    e^{\frac{i\pi}{2}\left(-M_{5i}+M_{6i}+M_{\ti{5}i}-M_{\ti{6}i}\right)\alpha_0^i}
    e^{\frac{i}{2}\left(-\phi^5+\phi^6+\tp^{\ti{5}}-\tp^{\ti{6}}\right)},\\
\kern -2em S^{+\dot{-}}&=&i e^{\frac{i\pi}{8}\left(\begin{smallmatrix}
    +M_{55} & - M_{65} & - 3M_{\ti{5} 5} & + 3M_{\ti{6} 5} \\
    +3M_{56} & + M_{66} & + 3M_{\ti{5} 6} & - 3M_{\ti{6} 6} \\
    +5M_{5\ti{5}} & - 5M_{6\ti{5}} & - 3M_{\ti{5} \ti{5}} & + 3M_{\ti{6} \ti{5}} \\
    -5M_{5\ti{6}} & + 5M_{6\ti{6}} & - M_{\ti{5} \ti{6}} & - 3M_{\ti{6} \ti{6}} \\
    \end{smallmatrix}
    \right)}
    e^{\frac{i\pi}{2}\left(M_{5i}-M_{6i}-M_{\ti{5}i}+M_{\ti{6}i}\right)\alpha_0^i}
    e^{\frac{i}{2}\left(\phi^5-\phi^6-\tp^{\ti{5}}+\tp^{\ti{6}}\right)},\\
\kern -2em S^{-\dot{-}}&=& i e^{\frac{i\pi}{8}\left(\begin{smallmatrix}
    -3M_{55} & + 3M_{65} & + M_{\ti{5} 5} & - M_{\ti{6} 5} \\
    -M_{56} & - 3M_{66} & - M_{\ti{5} 6} & + M_{\ti{6} 6} \\
    -3M_{5\ti{5}} & + 3M_{6\ti{5}} & - 3M_{\ti{5} \ti{5}} & + 3M_{\ti{6} \ti{5}} \\
    +3M_{5\ti{6}} & - 3M_{6\ti{6}} & - M_{\ti{5} \ti{6}} & - 3M_{\ti{6} \ti{6}} \\
    \end{smallmatrix}
    \right)}
    e^{\frac{i\pi}{2}(-M_{5i}+M_{6i}-M_{\ti{5}i}+M_{\ti{6}i})\alpha_0^i}
    e^{\frac{i}{2}(-\phi^5+\phi^6-\tp^{\ti{5}}+\tp^{\ti{6}})},
\eea
\bea
\kern -2em S^{+\dot{1}}&=&ie^{\frac{i\pi}{8}\left(\begin{smallmatrix}
+M_{55} & -M_{65} & +3M_{\ti{5}5} & +3M_{\ti{6}5} \\
+3M_{56} & + M_{66} & - 3M_{\ti{5}6} & -3M_{\ti{6}6} \\
-M_{5\ti{5}} & +M_{6\ti{5}} & -3M_{\ti{5}\ti{5}}& +M_{\ti{6}\ti{5}} \\
-M_{5\ti{6}}& +M_{6\ti{6}} & -3M_{\ti{5}\ti{6}} & +M_{\ti{6}\ti{6}} \end{smallmatrix}\right)}
e^{\frac{i\pi}{2} \left(M_{5i}-M_{6i}-M_{\ti{5}i}-M_{\ti{6}i}\right)\alpha_0^i} e^{\frac{i}{2}\left(\phi^5-\phi^6-\tp^{\ti{5}}-\tp^{\ti{6}}\right)}, \\
\kern -2em S^{- \dot{1}}&=&ie^{\frac{i\pi}{8}\left(\begin{smallmatrix}
-3M_{55} & + 3M_{65} & -M_{\ti{5}5} & -M_{\ti{6}5} \\
-M_{56} & - 3M_{66} & + M_{\ti{5}6} & +M_{\ti{6}6} \\
-M_{5\ti{5}} & + M_{6\ti{5}} & -3M_{\ti{5}\ti{5}}& +M_{\ti{6}\ti{5}} \\
-M_{5\ti{6}}& + M_{6\ti{6}} & -3M_{\ti{5}\ti{6}} & +M_{\ti{6}\ti{6}} \end{smallmatrix}\right)}
e^{\frac{i\pi}{2}\left(-M_{5i}+M_{6i}-M_{\ti{5}i}-M_{\ti{6}i}\right)\alpha_0^i}
e^{\frac{i}{2}\left(-\phi^5+\phi^6-\tp^{\ti{5}}-\tp^{\ti{6}}\right)}, \\
\kern -2em S^{+\dot{2}}&=&ie^{\frac{i\pi}{8}\left(\begin{smallmatrix}
+M_{55} & -M_{65} & +3M_{\ti{5}5} & +3M_{\ti{6}5} \\
+3M_{56} & + M_{66} & - 3M_{\ti{5}6} & -3M_{\ti{6}6} \\
-5M_{5\ti{5}} & +5M_{6\ti{5}} & + M_{\ti{5}\ti{5}}& -3M_{\ti{6}\ti{5}} \\
-5M_{5\ti{6}}& +5M_{6\ti{6}} & + M_{\ti{5}\ti{6}} & -3M_{\ti{6}\ti{6}} \end{smallmatrix}\right)}
e^{\frac{i\pi}{2}\left(M_{5i}-M_{6i}+M_{\ti{5}i}+M_{\ti{6}i}\right)\alpha_0^i}
e^{\frac{i}{2}\left(\phi^5-\phi^6+\tp^{\ti{5}}+\tp^{\ti{6}}\right)}, \\
\kern -2em S^{-\dot{2}}&=&ie^{\frac{i\pi}{8}\left(\begin{smallmatrix}
-3M_{55} & + 3M_{65} & - M_{\ti{5}5} & - M_{\ti{6}5} \\
-M_{56} & - 3M_{66} & + M_{\ti{5}6} & + M_{\ti{6}6} \\
+3M_{5\ti{5}} & - 3M_{6\ti{5}} & + M_{\ti{5}\ti{5}}& - 3M_{\ti{6}\ti{5}} \\
+3M_{5\ti{6}}& - 3M_{6\ti{6}} & + M_{\ti{5}\ti{6}} & - 3M_{\ti{6}\ti{6}} \end{smallmatrix}\right)}
e^{\frac{i\pi}{2}\left(-M_{5i}+M_{6i}+M_{\ti{5}i}+M_{\ti{6}i}\right)\alpha_0^i}
e^{\frac{i}{2}\left(-\phi^5+\phi^6+\tp^{\ti{5}}+\tp^{\ti{6}}\right)},
\eea
\bea
\kern -2em S^{\dot{1}\dot{+}}&=&-ie^{\frac{i\pi}{8}\left(\begin{smallmatrix}
-3M_{55} & + M_{65} & - M_{\ti{5}5} & + M_{\ti{6}5} \\
-3M_{56} & + M_{66} & - M_{\ti{5}6} & + M_{\ti{6}6} \\
+3M_{5\ti{5}} & + 3M_{6\ti{5}} & + M_{\ti{5}\ti{5}}& - M_{\ti{6}\ti{5}} \\
-3M_{5\ti{6}}& - 3M_{6\ti{6}} & + 3M_{\ti{5}\ti{6}} & + M_{\ti{6}\ti{6}} \end{smallmatrix}\right)}
e^{\frac{i\pi}{2} \left(-M_{5i}-M_{6i}+M_{\ti{5}i}-M_{\ti{6}i}\right)\alpha_0^i} e^{\frac{i}{2}\left(-\phi^5-\phi^6+\tp^{\ti{5}}-\tp^{\ti{6}}\right)},\\
\kern -2em S^{\dot{2}\dot{+}}&=&-ie^{\frac{i\pi}{8}\left(\begin{smallmatrix}
+M_{55} & - 3M_{65} & - 5M_{\ti{5}5} & + 5M_{\ti{6}5} \\
+M_{56} & - 3M_{66} & - 5M_{\ti{5}6} & + 5M_{\ti{6}6} \\
+3M_{5\ti{5}} & + 3M_{6\ti{5}} & + M_{\ti{5}\ti{5}}& - M_{\ti{6}\ti{5}} \\
-3M_{5\ti{6}}& - 3M_{6\ti{6}} & + 3M_{\ti{5}\ti{6}} & + M_{\ti{6}\ti{6}} \end{smallmatrix}\right)}
e^{\frac{i\pi}{2}\left(M_{5i}+M_{6i}+M_{\ti{5}i}-M_{\ti{6}i}\right)\alpha_0^i}
e^{\frac{i}{2}\left(\phi^5+\phi^6+\tp^{\ti{5}}-\tp^{\ti{6}}\right)}, \\
\kern -2em S^{\dot{1}\dot{-}}&=&-ie^{\frac{i\pi}{8}\left(\begin{smallmatrix}
-3M_{55} & + M_{65} & - M_{\ti{5}5} & + M_{\ti{6}5} \\
-3M_{56} & + M_{66} & - M_{\ti{5}6} & + M_{\ti{6}6} \\
-M_{5\ti{5}} & - M_{6\ti{5}} & - 3M_{\ti{5}\ti{5}}& + 3M_{\ti{6}\ti{5}} \\
+M_{5\ti{6}}& + M_{6\ti{6}} & - M_{\ti{5}\ti{6}} & - 3M_{\ti{6}\ti{6}} \end{smallmatrix}\right)}
e^{\frac{i\pi}{2}\left(-M_{5i}-M_{6i}-M_{\ti{5}i}+M_{\ti{6}i}\right)\alpha_0^i}
e^{\frac{i}{2}\left(-\phi^5-\phi^6-\tp^{\ti{5}}+\tp^{\ti{6}}\right)}, \\
\kern -2em S^{\dot{2}\dot{-}}&=&-ie^{\frac{i\pi}{8}\left(\begin{smallmatrix}
+M_{55} & - 3M_{65} & + 3M_{\ti{5}5} & - 3M_{\ti{6}5} \\
+M_{56} & - 3M_{66} & + 3M_{\ti{5}6} & - 3M_{\ti{6}6} \\
-M_{5\ti{5}} & - M_{6\ti{5}} & - 3M_{\ti{5}\ti{5}}& + 3M_{\ti{6}\ti{5}} \\
+M_{5\ti{6}}& + M_{6\ti{6}} & - M_{\ti{5}\ti{6}} & - 3M_{\ti{6}\ti{6}} \end{smallmatrix}\right)}
e^{\frac{i\pi}{2}\left(M_{5i}+M_{6i}-M_{\ti{5}i}+M_{\ti{6}i}\right)\alpha_0^i}
e^{\frac{i}{2}\left(\phi^5+\phi^6-\tp^{\ti{5}}+\tp^{\ti{6}}\right)},
\eea
\bea
\kern -2em S^{\dot{1}\dot{1}}&=&-ie^{\frac{i\pi}{8}\left(\begin{smallmatrix}
-3M_{55} & + M_{65} & - 3M_{\ti{5}5} & - 3M_{\ti{6}5} \\
-3M_{56} & + M_{66} & - 3M_{\ti{5}6} & - 3M_{\ti{6}6} \\
+M_{5\ti{5}} & + M_{6\ti{5}} & - 3M_{\ti{5}\ti{5}}& + M_{\ti{6}\ti{5}} \\
+M_{5\ti{6}}& + M_{6\ti{6}} & - 3M_{\ti{5}\ti{6}} & + M_{\ti{6}\ti{6}} \end{smallmatrix}\right)}
e^{\frac{i\pi}{2} \left(-M_{5i}-M_{6i}-M_{\ti{5}i}-M_{\ti{6}i}\right)\alpha_0^i} e^{\frac{i}{2}\left(-\phi^5-\phi^6-\tp^{\ti{5}}-\tp^{\ti{6}}\right)},\\
\kern -2em S^{\dot{2}\dot{1}}&=&-ie^{\frac{i\pi}{8}\left(\begin{smallmatrix}
+M_{55} & - 3M_{65} & + M_{\ti{5}5} & + M_{\ti{6}5} \\
+M_{56} & - 3M_{66} & + M_{\ti{5}6} & + M_{\ti{6}6} \\
+M_{5\ti{5}} & + M_{6\ti{5}} & - 3M_{\ti{5}\ti{5}}& + M_{\ti{6}\ti{5}} \\
+M_{5\ti{6}}& + M_{6\ti{6}} & - 3M_{\ti{5}\ti{6}} & + M_{\ti{6}\ti{6}} \end{smallmatrix}\right)}
e^{\frac{i\pi}{2}\left(M_{5i}+M_{6i}-M_{\ti{5}i}-M_{\ti{6}i}\right)\alpha_0^i}
e^{\frac{i}{2}\left(\phi^5+\phi^6-\tp^{\ti{5}}-\tp^{\ti{6}}\right)}, \\
\kern -2em S^{\dot{1}\dot{2}}&=&-ie^{\frac{i\pi}{8}\left(\begin{smallmatrix}
-3M_{55} & + M_{65} & - 3M_{\ti{5}5} & - 3M_{\ti{6}5} \\
-3M_{56} & + M_{66} & - 3M_{\ti{5}6} & - 3M_{\ti{6}6} \\
+5M_{5\ti{5}} & + 5M_{6\ti{5}} & + M_{\ti{5}\ti{5}}& - 3M_{\ti{6}\ti{5}} \\
+5M_{5\ti{6}}& + 5M_{6\ti{6}} & + M_{\ti{5}\ti{6}} & - 3M_{\ti{6}\ti{6}} \end{smallmatrix}\right)}
e^{\frac{i\pi}{2}\left(-M_{5i}-M_{6i}+M_{\ti{5}i}+M_{\ti{6}i}\right)\alpha_0^i}
e^{\frac{i}{2}\left(-\phi^5-\phi^6+\tp^{\ti{5}}+\tp^{\ti{6}}\right)}, \\
\kern -2em S^{\dot{2}\dot{2}}&=&-ie^{\frac{i\pi}{8}\left(\begin{smallmatrix}
+M_{55} & - 3M_{65} & + M_{\ti{5}5} & + M_{\ti{6}5} \\
+M_{56} & - 3M_{66} & + M_{\ti{5}6} & + M_{\ti{6}6} \\
-3M_{5\ti{5}} & - 3M_{6\ti{5}} & + M_{\ti{5}\ti{5}}& - 3M_{\ti{6}\ti{5}} \\
-3M_{5\ti{6}}& - 3M_{6\ti{6}} & + M_{\ti{5}\ti{6}} & - 3M_{\ti{6}\ti{6}} \end{smallmatrix}\right)}
e^{\frac{i\pi}{2}\left(M_{5i}+M_{6i}+M_{\ti{5}i}+M_{\ti{6}i}\right)\alpha_0^i}
e^{\frac{i}{2}\left(\phi^5+\phi^6+\tp^{\ti{5}}+\tp^{\ti{6}}\right)}.\label{newspinend}
\eea
\section{Three-point function with twist \texorpdfstring{$n$}{} spin fields}
\label{app_3pf}
In section \ref{3pf_sigma2} we computed the three-point function of a pair of twist 2 spin fields and an untwisted fermion field and discussed the appropriate alignment of branch cuts. In this appendix we generalize this computation to a three-point function which contains a pair of spin fields at general twists $n$ order (where $n$ is even), as well as an untwisted fermion. We use the map
\be
\frac{z-z_1}{z-z_2}=\left(a\frac{t-t_1}{t-t_2}\right)^n,
\ee
and consider a non-twist operator at location $z_0$.  The above obviously has $n$ distinct solutions for $z=z_0$, given by
\be
t_{\omega_k}=\frac{\omega_k\frac{(z_0-z_1)^{1/n}}{(z_0-z_2)^{1/n}}-at_1}{\omega_k\frac{(z_0-z_1)^{1/n}}{(z_0-z_2)^{1/n}}-a},
\ee
where $k\in\{0,\cdots,n-1\}$, and
\be
\omega_k=e^{2\pi i k/n}
\ee
are the $n^{\rm th}$ roots of unity.  Using this, one can find the expression
\be
\left.\frac{dz}{dt}\right|_{t_{\omega_k}}=n\frac{z_{01}z_{02}}{z_{12}}\frac{t_{12}} {t_{\omega_k 1}t_{\omega_{k 2}}},
\ee
where $t_{\omega_k 1}=t_{\omega_k}-t_1$.  The lift of a fermion in the first patch $\psi^{\alpha \dot{A}}_1$ lifts to the image
\be
\psi^{\alpha \dot{A}}_1(z_0)\rightarrow \frac{1}{\sqrt{n}}
\frac{\sqrt{z_{12}}}{\sqrt{z_{01}}\sqrt{z_{02}}}
\frac{\sqrt{t_{\omega_{k 1}}} \sqrt{t_{\omega_k 2}}}{\sqrt{t_{12}}}\psi(t_{\omega_k}),
\ee
where we pick the appropriate $\omega_k$ for the first patch.  We may easily note the differences between odd and even twist by tracking $z_{01}\rightarrow e^{2\pi i} z_{01}$ which results in $t_{\omega_k}\rightarrow t_{\omega_{k+1}}$ for $k< n-1$, and $t_{\omega_{(n-1)+1}}=e^{2\pi i}t_{\omega_0}$.

We now use this to construct a three point function with two twist $n$ (assumed even) operators and a non-twist sector bi-fermion.  Again, we start with the operator
\be
\psi^{\alpha \dot{A}}_1\psi^{\beta \dot{B}}_2
\ee
and consider the images.  There are a total of $\binom{N}{2}\times 2!=N(N-2)$ individual terms (recall, the order of $1,2$ matters because we have fermions).  Next, when considering such an operator, we must consider the full $S_N$ invariant, and so combinations of both $\psi^{\alpha \dot{A}}_1\psi^{\beta \dot{B}}_2$ and $\psi^{\alpha \dot{A}}_2\psi^{\beta \dot{B}}_1$ will show up, again restricting the combination of superscripts to be antisymmetric.  When such an operator is placed in a correlator with two twist $n$ insertions, the cycle $(1,2,3...n)$ must appear with its inverse for the boundary conditions to be satisfiable.  Choosing such a representative term, one must keep only those operators of the form $\psi_i \psi_j$ where $i$ and $j$ both appear in the cycle.  If one of the terms, for example $\psi_j$, appears with $j$ not in the set $\{1,2...,n\}$, then there will be a bare expectation value of this fermion, which is zero.


Given all of this, we can consider the operator
\bea
{\mathcal{O}}_0^{++}=\epsilon_{\dot{A}\dot{B}}\Bigg(\psi^{+ \dot{A}}_1 \psi^{+ \dot{B}}_2\!\!\!&+&\!\!\!\psi^{+ \dot{A}}_1 \psi^{+ \dot{B}}_3+\psi^{+ \dot{A}}_1 \psi^{+ \dot{B}}_4+\cdots+\\
 &+&\!\!\!\psi^{+ \dot{A}}_2 \psi^{+ \dot{B}}_3+\psi^{+ \dot{A}}_2 \psi^{+ \dot{B}}_4+\cdots+\nn\\
 &&\qquad\quad\;+\;{\psi^{+ \dot{A}}_3 \psi^{+ \dot{B}}_4} + \ddots \qquad\Bigg)\nn \\
 &&\!\!\!\!\!\!\!\!\!\!\!\!\!\!\!\!\!\!\!\!\!\!\!\!\!\!\!\!\!\!\!\!\!\!\!\!\!\!\!\!\!\!\!\!\!\!\!\!\!\!\!\!\!\!
 =\frac{1}{2}\sum_{p,q, p\neq q}\epsilon_{\dot{A}\dot{B}}\psi^{+ \dot{A}}_{p+1} \psi^{+ \dot{B}}_{q+1}\,,\nn
\eea
where $p$ and $q$ run from $0$ to $n-1$.  Note that the copy label is $p+1$ because $p$ starts at 0.  Thus, we consider
\be
A_{3,n}=\langle {\mathcal{O}}_0^{++}(z_0) \sigma^{-\dot{+}}_{(1,2,3,...,n)} \sigma^{-\dot{-}}_{(n,n-1,n-2,...,1)}\rangle
\ee
as our representative calculation.  Of course, there will be some combinatoric factors that come out when considering the full amplitude, but we do not wish to complicate our discussion here.  Lifting to the cover, we find
\bea
A_{3,n}\rightarrow \frac{1}{2}\sum_{p,q, p\neq q}\epsilon_{\dot{A}\dot{B}}&& \kern -1.5em \Biggl\langle\frac{1}{\sqrt{n}} \frac{\sqrt{z_{12}}}{\sqrt{z_{01}}\sqrt{z_{02}}}
\frac{\sqrt{t_{(\omega_p 1)}} \sqrt{t_{(\omega_p 2)}}}{\sqrt{t_{12}}}\psi^{+\dot{A}}(t_{\omega_p}) \nn \\
&& \qquad \quad \times \frac{1}{\sqrt{n}}\frac{\sqrt{z_{12}}}{\sqrt{z_{01}}\sqrt{z_{02}}} \frac{\sqrt{t_{(\omega_q 1)}} \sqrt{t_{(\omega_q 2)}}}{\sqrt{t_{12}}}\psi^{+\dot{B}}(t_{\omega_{q}})\nn \\
&& \qquad \quad \times |b_1|^{-1/(2n)} S^{-\dot{+}}_1 |b_2|^{-1/(2n)}S^{-\dot{-}}_2\Biggr\rangle.
\eea
This calculation is now almost identical to the last one.  We only have to contend with summing over the various points and assign signs.  We give a quick example of this for $n=4$ for the cycle $(1,2,3,4)$.  In this case, there are 6 operators showing up with subscripts
\be
(12), (13), (14), (23), (24), (34).
\ee
Adjacent patches are assigned a $-1$ while gapped (by 2) are assigned a $+1$.  This gives
\be
-1, +1, -1, -1, +1, -1,
\ee
and adding these gives $-2=-4/2$.  If we had done this for $n=6$ we would find
\be
(12), (13), (14), (15), (16), (23), (24), (25), (26), (34), (35), (36), (45), (46), (56),
\ee
we get
\be
-1,1,-1,1,-1,-1,1,-1,1,-1,1,-1,-1,1,-1
\ee
and upon adding, we get $-3=-6/2$.  We will generalize this result below to $-n/2$.

We note that if the two image patches are adjacent, for example $q=p+1$, then we get a minus sign when continuously deforming from patch one.  This in fact happens for all $q=p+\ell$ where $\ell$ is odd.  Similarly, we will get positive signs when $q=p+\ell$ when $\ell$ is even.  We count the number of images that have adjacent patch combinations: $12, 23, 34, ...,(n-1)n, n1$.  Thus, there are $n$ such terms.  We may again count the number of next to adjacent: $(13,24,35,... (n-2)n,(n-1)1,n2)$ and so we again find $n$ such terms.  The plus signs from before the adjacent cancel the signs from the next to adjacent.  This continues to happen until we look at $\ell=n/2$, i.e. skipping by half the patches.  In this case, there are only $n/2$ terms: we do not count the pair $(1,1+n/2)$ and $(1+n/2,1+n/2+n/2)=(1+n/2,1)$ as distinct.  In the $n=4$ case above, note that $(1,4)$ appears, but not $(4,1)$.

So, is the overall contribution positive or negative?  We break into two cases: $n/2$ odd, and $n/2$ even.  If $n/2$ is odd, then the contributions from this last step are $n/2$ minus signs.  It is also true in this case that $n/2-1$ is even, and so in the earlier steps, there were just as many positive signs as negative, and so these cancel.  Thus, the overall contribution is $-n/2$.  In the case where $n/2$ is even, then the final step gives $n/2$ plus signs.  However, now $n/2-1$ is odd, which means there is one more negative contribution than positive, and all others cancel.  This gives a contribution of $n$ negative signs, along with the last $n/2$ positive signs, resulting in an overall contribution of $-n/2$.  Hence, the answer is always $-n/2$.  This factor of $n$ in the numerator cancels that coming from the $1/\sqrt{n}^2$ from the conformal dimension part of the transformation of $\psi$ operators.

Putting this all in, we get
\bea
&&A_{3,n}\rightarrow \frac{z_{12}}{z_{01}z_{02}|t_{12}|}|b_1|^{-1/(2n)} |b_2|^{-1/(2n)},
\eea
which is slightly different because the powers of the $b_i$ are different, and also the $b_i$ are now defined by
\be
z-z_1=a^n\frac{z_{12}}{(t_{12})^n}(t-t_1)^n+\cdots,\qquad z-z_2=-\frac{1}{a^n}\frac{ z_{12}}{(t_{12})^n}(t-t_2)^n+\cdots.
\ee
Putting this together, we find
\be
\langle {\mathcal{O}}_0^{++}(z_0) \sigma^{-+}_{(1,2,3,...,n)} \sigma^{--}_{(n,n-1,n-2,...,1)}\rangle\rightarrow \frac{z_{12}}{z_{01}z_{02}|z_{12}|^{1/n}},
\ee
agreeing with the $n=2$ case from earlier.  Writing the strict equality is now done using a different conformal map, and so the conformal anomaly is slightly different.  We again assume normalized twists, and the contribution from the conformal anomaly is given by $|z_{12}|^{-n+1/n}$, resulting in
\be
\langle {\mathcal{O}}_0^{++}(z_0) \sigma^{-+}_{(1,2,3,...,n)} \sigma^{--}_{(n,n-1,n-2,...,1)}\rangle=\frac{z_{12}}{z_{01}z_{02}|z_{12}|^{n}},
\ee
which agrees with the form of a three point function for operators of weights $(1,0)$, $(n/4,n/4)$ and $(n/4,n/4)$.

We generalize this result to
\be
\langle {\mathcal{O}}^{\alpha {\dot{A}} \beta \dot{B}}(z_0) \sigma^{W\ti{X}}_{(1,2,3,...,n)} \sigma^{Y\ti{Z}}_{(n,n-1,n-2,...,1)}\rangle=\left(1/2[\Gamma^{\alpha\dot{A}},\Gamma^{\beta \dot{B}}]C\right)^{W\ti{X}Y\ti{Z}}\frac{z_{12}}{z_{01}z_{02}|z_{12}|^{n}},
\ee
where
\be
\mathcal{O}^{\alpha {\dot{A}} \beta \dot{B}}=\frac{1}{2}\sum_{p,q,p\neq q}2\psi_{p+1}^{\left[\alpha \dot{A}\right.}\psi_{q+1}^{\left.\beta \dot{B}\right]}.
\ee
Note that switching the second and third twist operators (indices and positions) results in the same correlator, owing to the antisymmetry of the matrix $[\Gamma^{\alpha\dot{A}},\Gamma^{\beta \dot{B}}]C$, which would not be possible without the cocycles.

\end{document}